\documentclass[twocolumn]{aastex63}

\newcommand{\str}{Str\"{o}mgren}
\newcommand{\ars}{\langle R_{\rm s} \rangle}

\newcommand{\nH}{n_{\rm H,\infty}}
\newcommand{\cmq}{{\rm cm}^{-3}}

\newcommand{\hii}{\rm H\,{\textsc{ii}}}

\newcommand{\msun}{M_{\sun}}

\newcommand{\mbh}{M_{\rm BH}}

\newcommand{\khp}[1]{{\color{black}#1}}
\newcommand{\khpii}[1]{{\color{black}#1}}

\submitjournal{ApJ}

\shorttitle{Biconical-dominated accretion flow in a hyperaccretion}
\shortauthors{Park et al.}

\begin{document}

\title{Biconical-dominated Accretion Flow onto Seed Black Holes in a Hyperaccretion Regime}

\correspondingauthor{KwangHo Park}
\email{kwangho.park@physics.gatech.edu}

\author[0000-0001-7973-5744]{KwangHo Park}
\affiliation{School of Physics, Georgia Institute of Technology, Atlanta,
GA 30332, USA}

\author[0000-0003-1173-8847]{John H. Wise}
\affiliation{School of Physics, Georgia Institute of Technology, Atlanta,
GA 30332, USA}

\author[0000-0002-7835-7814]{Tamara Bogdanovi\'c}
\affiliation{School of Physics, Georgia Institute of Technology, Atlanta,
GA 30332, USA}

\author[0000-0003-4223-7324]{Massimo Ricotti}
\affiliation{Department of Astronomy, University of Maryland, College Park, MD 20740, USA}



\begin{abstract}

Hyperaccretion occurs when the gas inflow rate onto a black hole (BH) is so high that the radiative feedback cannot reverse the accretion flow. This extreme process is a promising mechanism for the rapid growth of seed BHs in the early universe, which can explain high-redshift quasars powered by billion solar mass BHs. In theoretical models, spherical symmetry is commonly adopted for \khp{hyperaccretion flows; however the sustainability of such structures} on timescales corresponding to the BH growth has not been addressed yet. Here we show that stochastic interactions between the ionizing radiation from the BH and nonuniform accretion flow can lead to \khp{the formation of a rotating gas disk} around the BH. \khp{Once the disk forms}, the supply of gas to the BH preferentially occurs via biconical-dominated accretion flow perpendicular to the disk, avoiding the centrifugal barrier of the disk. \khp{Biconical-dominated accretion flows} from opposite directions collide in the vicinity of the BH supplying high-density, low angular momentum gas to the BH, whereas most of the gas with nonnegligible angular momentum is deflected to the rotationally supported outflowing decretion disk. The disk becomes reinforced progressively as more mass from the biconical flow transfers to the disk and some of the outflowing gas from the disk is redirected to the biconical accretion funnels through a meridional structure. This axisymmetric hydrodynamic structure of a \khp{biconical-dominated accretion flow} and decretion disk continues to provide  uninterrupted flow of high-density gas to the BH.
\end{abstract}




\section{Introduction}

Observations of the high redshift universe have uncovered quasars being powered by supermassive black holes (SMBHs) with masses $\gtrsim 10^9\,\msun$ \citep[e.g.,][]{Fan:2001,Willott:2003,Mortlock:2011,Banados:2018}. The estimated mass range is comparable to the most massive black holes (BHs) in the local universe, and thus it raises challenging questions about the origin and growth history of massive black holes. 

There are three scenarios that explain the origins of BH seeds in the mass range of $10^2$--$10^5\,\msun$, i.e, intermediate-mass BHs. They suggest that special conditions in the early universe can produce seed BHs more massive than typical stellar remnants as the products of the deaths of very massive metal-free (Population\,III) stars \citep{BrommCL:99,AbelBN:00,MadauR:01}, \khp{a dense stellar cluster collapse \citep{Devecchi:2009, Davies:2011, Lupi:2014, Katz:2015,Boekholt:2018, Reinoso:2018}}, and a direct collapse of chemically pristine gas \citep{BegelmanVR:06, ChoiSB:13, YueFSXC:14, Regan:2017}. At the Eddington limit, these BH seeds can grow by $\sim$7 orders of magnitude over 700\,Myr. That is approximately 20 $e$-folding times, given a radiative efficiency of $\eta = 0.1$ in a thin disk model \citep{ShakuraS:73}. However, it is still not clearly understood how a BH can sustain an extremely high accretion rate for such a long period. Therefore, the high-mass end scenario ($\sim 10^5\,\msun$) for seeds such as a direct collapse BH (DCBH) is preferred\khp{. However}, even DCBHs still must go through a rapid growth period to build up mass by $\gtrsim 4$ orders of magnitude to account for the earliest SMBHs at $z \gtrsim 6$.

The biggest challenge in understanding the rapid growth of the seed BHs is the radiative feedback from the BHs themselves. \khp{Abundant neutral gas} at $T\sim 10^4\,$K in the early universe might provide ideal conditions for achieving a high accretion rate, but several numerical studies find that the radiative feedback from BHs efficiently suppresses growth \citep{MiloCB:09, ParkR:11, ParkR:12, ParkR:13, ParkRDR:14b, PacucciVF:2015}. When a BH accretes gas, the BH radiation heats and ionizes the ambient gas and forms a low-density hot bubble, which \khp{limits} gas accretion. The causal relationship of accretion and the resulting radiation creates a feedback loop between the two, ultimately leading to an oscillatory behavior of accretion rate.
As a result, even though cold gas is abundant in the early universe, the mean accretion rate is only $\sim$1\% of the Bondi rate \citep{Bondi:52}, which makes it almost impossible even for DCBHs to grow to high-redshift quasars \citep{ParkR:11, ParkR:12, Aykutalp:2014}. 
\citet*[][hereafter PWB17]{ParkWB:2017} showed consistent results using 3-dimensional (3D) radiation-hydrodynamic simulations. They emphasized that the nature of the accretion flow is turbulent due to the constantly changing ionized region, but the quasi-spherical symmetry is maintained as the thermal energy of the gas remains dominant over the turbulent kinetic energy. \khp{We note that \citet{Regan:2019} show that mechanical feedback from a seed BH is not effective in regulating gas supply but still can suppress the accretion rate to 0.1--0.5 times the Eddington rate.}

When an ionized sphere (i.e., \khp{a} \str~sphere) forms, the ionization front (I-front) becomes subject to Rayleigh-Taylor instabilities, as the buoyant ionized gas rises into a dense neutral medium that is infalling from the gravitational pull of the BH. However, Rayleigh-Taylor instabilities across the I-front are suppressed efficiently by the BH radiation \citep{ParkRDR:14a, Ricotti:2014}.
Interestingly, \citet{ParkRDR:14a} also shows that the stability of the I-front is determined by the relative scale of the \str~sphere ($R_{\rm s}$) and the Bondi radius ($r_{\rm B}$). When $R_{\rm s} < r_{\rm B}$, neutral gas can \khp{accrete to} the BH without being interrupted by the ionizing radiation. This criterion \khp{that is} determined at a relatively large scale provides a condition for hyperaccretion \citep{Begelman:79, Begelman:2012a, Inayoshi:2015a, Sakurai:2016}, where the ionizing radiation from the BH cannot regulate the gas accretion as ionizing photons are trapped within an accretion flow. Assuming spherical symmetry, the radiation-regulated accretion onto BHs becomes similar to the classical Bondi accretion problem since the radiation does not play any significant role in regulating gas accretion \citep{PacucciF:2015, PacucciVF:2015}. In this so-called {\it feeding-dominated} regime, the Bondi rate can easily reach up to $\sim$1000 times the Eddington rate given \khp{a} dense environment around the seed BHs. \khpii{For example, this scenario corresponds to a light seed BH trapped in a stellar cluster being fed by dense gas \citep{AlexanderN:2014} or a BH embedded in a stellar bulge, \khpii{which} can accrete at super-Eddington rates \citep{Park:2016}.}

It is common to assume spherical symmetry for hyperaccretion \khp{flows}; however, it is not clear how a deviation from spherical symmetry would operate. Axisymmetric hydrodynamic or magneto-hydrodynamic simulations of Bondi accretion show that low angular momentum gas forms a dense torus around a BH, significantly reducing the accretion rate \citep{Abramowicz:1981,ProgaB:2003a, ProgaB:2003b}. However, some studies show that a deviation from spherical symmetry still can lead to a hyperaccretion state. For example, when the BH radiation displays a preferred direction (i.e., perpendicular to the accretion disk), the gas is pushed away from the BH in \khp{a} bipolar direction due to the radiative feedback, whereas continuous accretion can happen efficiently from the azimuthal direction \cite[e.g.,][]{Sugimura:2016, Takeo:2018}. However, the accretion from the azimuthal direction is easily stunted by the angular momentum barrier \citep{Sugimura:2018}, effectively blocking both effective accretion channels. 
 
In this study, we explore the 3D structure of hyperaccretion flows at the Bondi radius scale, focusing on their departure from spherical symmetry. We find that a biconical-dominated accretion flow (BDAF) develops after a stochastic interaction between the radiation and gas inflow. The resulting BDAFs from the opposite directions collide near the BH, maintaining a high density environment in the vicinity of the central BH. In Section~\ref{sec:method}, we describe our numerical simulation setup, and we present the results in Section~\ref{sec:results}. \khp{We discuss and summarize the results in Section~\ref{sec:discussion} and \ref{sec:summary}, respectively.}

\section{Methodology}
\label{sec:method}

\subsection{Radiation-hydrodynamic simulations}
We perform 3D radiation-hydrodynamic simulations to study the large-scale structure of hyperaccretion flows. We use the adaptive mesh refinement code {\it Enzo} equipped with the {\it Moray} package to evolve the radiative transfer equation \citep{Wise:2011,Bryan:2014,Brummel-Smith:2019}. 

\begin{table*}
   \begin{center}
   \caption{Simulation Parameters}
   \begin{tabular}{lcccccccc}
   \hline
   \hline
   ID   &  $\nH$  & $D_{\rm box}$  & $\Delta D_{\rm min}$  &  Spherical  & Remarks       \\
      &  $({\rm cm}^{-3})$  & (pc) & (pc) & Symmetry &         \\
   \hline
   {M4N3} &  $10^3$ & 40.0 &  0.156 & yes & Park et al. (2017) \\
   {M4N4} &  $10^4$ & 20.0 &  0.078 & yes & Park et al. (2017) \\
   {M4N5c} &  $10^5$ & 4.0 &  0.0156 & yes &  BH at $x=y=z=0.0078$\,pc \khp{=$\Delta D_{\rm min}/2$} \tablenotemark{$a$}\\
   {M4N5} &  $10^5$ & 4.0 &  0.0156 &  no &  BH at $x=y=z=0.004$\,pc \khp{$\approx\Delta D_{\rm min}/4$} \\
   {M4N5-free} &  $10^5$ & 4.0 &  0.0156 & no &   same as M4N5, BH position not fixed \\
   {M4N5c-free} &  $10^5$ & 4.0 &  0.0156 & no &   same as M4N5c, BH position not fixed \\
   {M4N5c-free-hi} &  $10^5$ & 4.0 &  0.0078 & no & high resolution \\
   {M4N6} &  $10^6$ & 4.0 &  0.0156 & no &  BH at $x=y=z=0.004\,$pc \khp{$\approx\Delta D_{\rm min}/4$}  \\    
   {M4N6-free-hi} &  $10^6$ & 4.0 &  0.0078 & no &    high resolution, BH position not fixed \\
   \hline
   \end{tabular}
   \tablecomments{$\mbh=10^4\,\msun$ for all runs. }
   \tablenotetext{a}{at the center of a cell of maximum refinement.}
   \label{table:para}
   \end{center}
   \end{table*}

We use the Eddington-limited Bondi recipe to calculate the BH accretion luminosity that sources the radiative transfer equation. The Bondi radius, defined to bound a region within which the BH gravitational potential dominates over the thermal energy of the gas, is
\begin{equation}
   r_{\rm B} = \frac{G\mbh}{c_\infty^2}
   \simeq 0.65\,{\rm pc} \left(\frac{\mbh}{10^4\,\msun}\right)
   \left(\frac{T_\infty}{8\times 10^3\,{\rm K}}\right)^{-1},
   \label{eq:bondiradius}
\end{equation}   
where $c_\infty$ is the sound speed of the gas outside the ionized region that is not affected by the BH radiation \khp{or gravity} and $T_\infty$ is the relevant gas temperature. We adopt $c_\infty = 9.1\,{\rm km\,s^{-1} (T_\infty /10^4\,K)^{1/2}}$ assuming isothermal gas with a mean molecular weight of 1. The Bondi accretion rate is defined as
\begin{equation}
   \dot{M}_{\rm B}=4\pi\,\lambda_{\rm B}\,r_{\rm B}^2\,\rho_\infty\,c_\infty ,
   \label{eq:bondiacc}
\end{equation}   
where the dimensionless accretion rate is $\lambda_{\rm B}=e^{3/2}/4 \approx 1.12$ for isothermal gas and $\rho_\infty$ is the gas density far from the BH. Since the gas is under the influence of the BH radiation, in order to calculate the BH accretion rate $\dot{M}_{\rm BH}$, we use the gas density $\rho_{\rm HII}$ and sound speed $c_{\rm HII}$ of the cell that contains the BH, arriving at
\begin{equation}
   \dot{M}_{\rm BH}=4\pi\,\lambda_{\rm B}\,r_{\rm B, HII}^2\,\rho_{\rm HII}\,c_{\rm HII},
   \label{eq:acc}
\end{equation}   
where $r_{\rm B, HII}=G\mbh/c_{\rm HII}^2$ is the accretion radius under the influence of radiation. The accretion radius for $T_{\rm HII}=4\times 10^4\,$K is $r_{\rm B, HII}=0.13
(\mbh/ 10^4\,\msun)\,$pc and \khp{is} well resolved with the typical resolution of the runs listed in Table~\ref{table:para} (i.e., $\Delta D_{\rm min}=0.0156\,$pc). 

The accretion rate is then converted into the BH luminosity as
\begin{equation}
   L_{\rm BH}= {\rm min} \left(\eta \dot{M}_{\rm BH} c^2, L_{\rm Edd}\right),
\end{equation}   
where we use the radiative efficiency of $\eta=0.1$ assuming a thin disk model \citep{ShakuraS:73} and $c$ is speed of light. Extending the thin disk model with a limited luminosity to the hyperaccretion regime \khp{is clearly an idealization but} can be justified as follows. Global 3D magneto-hydrodynamic simulations of super-Eddington accretion disks show that \khp{the BH luminosity converges to $\sim 10L_{\rm Edd}$ as} the radiative efficiency is $\sim$5\% for $\dot{M}_{\rm BH} = 220\,L_{\rm Edd}/c^2$ which is comparable to the standard thin disk model \citep{Jiang:2014} and drops to $\sim$1\% for $\dot{M}_{\rm BH}=1500\,L_{\rm Edd}/c^2$ \citep{Jiang:2019}. Here we apply the Eddington luminosity $L_{\rm Edd}$ as the upper limit that is
\begin{equation}
   L_{\rm Edd} = \frac{4\pi G\mbh m_{\rm p}c}{\sigma_{\rm T}} \simeq
   1.26\!\times\!10^{38}
   \left(\frac{\mbh}{\msun}
   \right)\,{\rm erg\ s}^{-1},
\end{equation}   
where $m_{\rm p}$ is the proton mass and $\sigma_{\rm T}$ is the Thomson cross-section for electrons. 

When hyperaccretion occurs, the accretion rate using Eq.~(\ref{eq:acc}), which works well in the feedback-limited regime, is unphysically high since $\rho_{\rm HII}$ does not represent the mean gas density within the radius $r_{\rm B, HII}$. Using Eq.~(\ref{eq:bondiacc}) one can \khp{instead} calculate the Bondi accretion rate that corresponds to $\sim 10^3$\,${\rm L}_{\rm Edd}/c^2$. Whether such \khp{a} high accretion rate onto the BH can be maintained in reality, all the way down to the event horizon, sensitively depends on the structure of the accretion flow. Here we show that the hyperaccretion accretion flow in general does not adhere to an idealized assumption of spherical symmetry but evolves to a geometry where only a small fraction of the solid angle allows uninterrupted inflow of gas into the BH. This implies that only a fraction of the mass accretion rate calculated in Eq.~(\ref{eq:bondiacc}) actually reaches the BH and that the rest must be expelled. Consequently, we limit the mass accretion rate onto the BH to a fiducial value that we assume is equal to $\dot{M}_{\rm Edd}=L_{\rm Edd}/(\eta c^2)$. Note that this is a simple choice that does not affect any of our results as long as $\dot{M}_{\rm BH} \gg \dot{M}_{\rm Edd}$ from Eq.~(\ref{eq:acc}), but it prevents a large fraction of gas from being lost from the computational domain.  In reality, the mass accretion rate may be higher than $L_{\rm Edd}/(\eta c^2)$, depending on the gas density and the structure of the accretion flow on scales close to the BH event horizon.
\begin{figure} 
   \includegraphics[width=\linewidth]{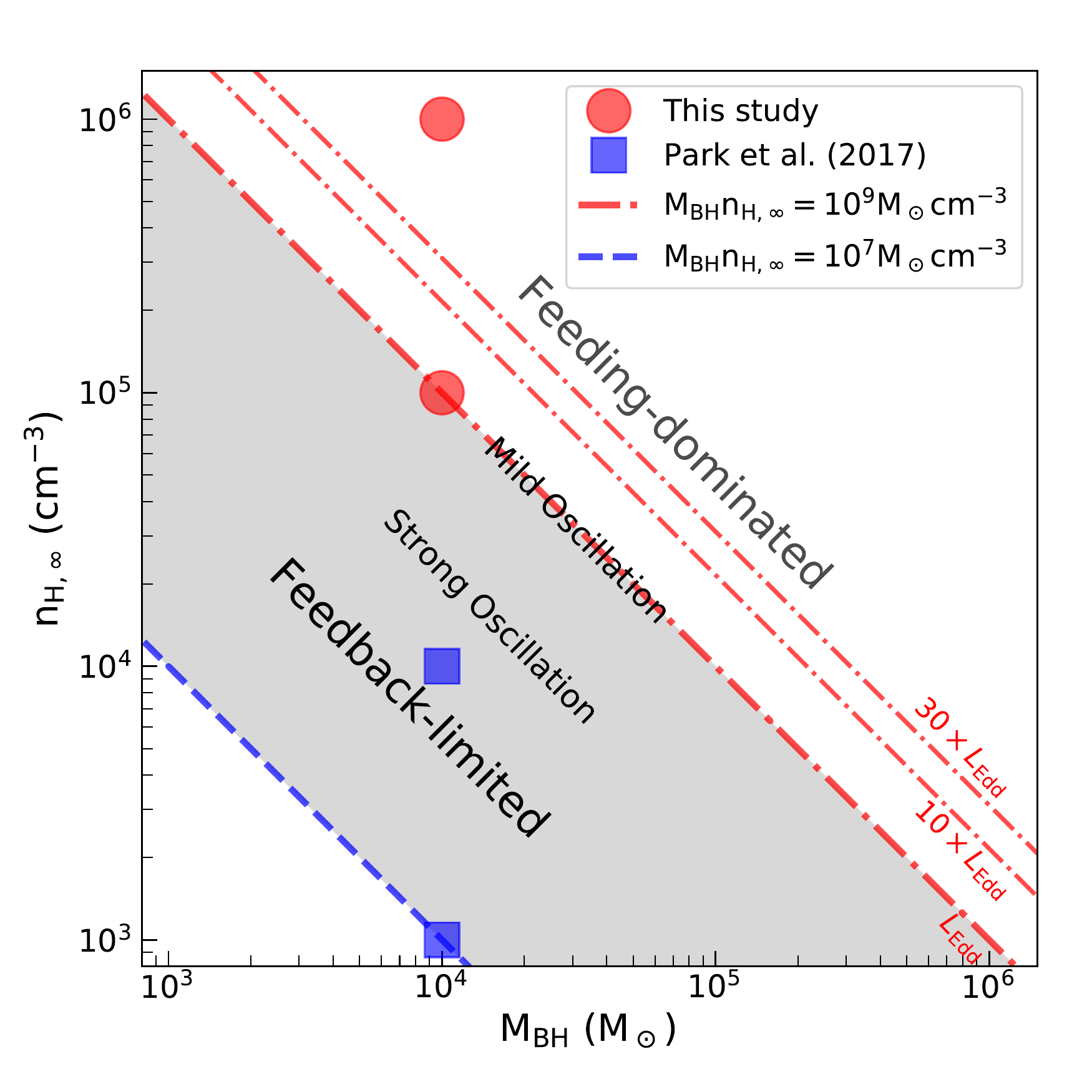}
   \caption{Accretion regimes as a function of $\mbh$ and
   $\nH$. Squares show the runs M4N3 and M4N4 in the feedback-limited
   regime from PWB17 and circles show the runs in the
   hyperaccretion regime $\mbh \nH \ga 10^9\,\msun \cmq$ in this
   study.} 
\label{fig:mn}
\end{figure}

Adopting the mean BH accretion rate in the feedback-dominated regime $\langle \dot{M}_{\rm BH} \rangle \sim 0.01 \dot{M}_{\rm B}$, the mean size of an ionized sphere is 
\begin{eqnarray}
   \langle&R_s&\rangle =
   \left( \frac{3 N_{\rm ion}}{4\pi \alpha_{\rm rec} \nH^{2}}  \right)^{\frac{1}{3}} 
   \sim 0.70\,{\rm pc} \nonumber \\
   &\times&\left(\frac{\mbh}{10^4\,\msun}\right)^{\frac{2}{3}}
   \left(\frac{\nH}{10^5\,{\rm cm}^{-3}}\right)^{-\frac{1}{3}}
   \left(\frac{T_\infty}{{\rm 8\times 10^3\,K}}\right)^{-\frac{1}{2}},
   \label{eq:Rs}
\end{eqnarray}      
where $N_{\rm ion}$ is the total number of ionizing photons in the range of $13.6\,$ eV $\le E \le 100$\,keV with a spectral index of $\alpha_{\rm spec}=1.5$ for power-law energy distribution, and $\alpha_{\rm rec}$ is the case B recombination coefficient. Comparing Equations~(\ref{eq:bondiradius}) and (\ref{eq:Rs}), the parameter space that we explore with $\mbh = 10^4\,\msun$ and $\nH = 10^5 - 10^6\,{\rm cm}^{-3}$ belongs to the hyperaccretion regime as $\langle R_s \rangle \lesssim  r_{\rm B}$ \citep{ParkRDR:14a}. 

Photo-ionization, photo-heating, and gas cooling are computed by the adaptive ray-tracing module {\it Moray} that couples these rates to the hydrodynamic equations (see section 2.2 of PWB17). Here, we use 4 energy bins of (28.4, 263.0, 2435.3, 22551.1) eV, each with a fractional luminosity of (0.6793, 0.2232, 0.0734, 0.0241), respectively (see section 2.3 in PWB17 for details).

\subsection{Transition from feedback-limited to feeding-dominated regime}

\begin{figure*}
   \includegraphics[width=\linewidth]{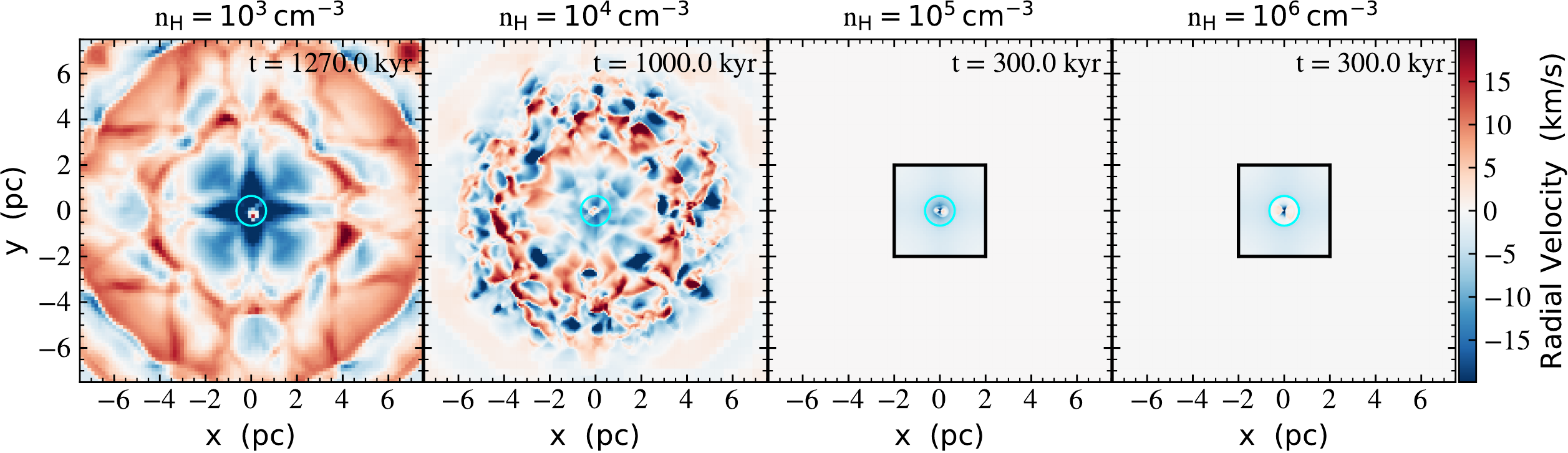}
   \caption{Radial velocity maps of M4N3 and M4N4 in the {\it feedback-limited} regime from PWB17 (left two panels) and M4N5 and M4N6 in the {\it feeding-dominated} regime (right two panels). Note the difference in the sizes of hydrodynamic structures in between two regimes. Cyan circles show the Bondi radius $r_{\rm B}$ and boxes show the simulation domain for M4N5 and M4N6.}
\label{fig:vr_comp} 
\end{figure*}

Fig.~\ref{fig:mn} shows distinct accretion regimes as a function of BH mass $\mbh$ and gas density $\nH$ assuming $\eta=0.1$ and $T_\infty=10^4\,$K. When $\mbh \nH < 10^9$\,$\msun \cmq$, the accretion is regulated by the radiative feedback from the BH. Squares in Fig.~\ref{fig:mn} show the runs from PWB17. In this regime, the size of the \str~sphere, ionized by the BH, is on average much larger than the Bondi radius. The hot bubble regulates the accretion onto the BH by blocking high-density gas. Applying similar setups used in PWB17, but with values $\mbh \nH \ge 10^9$\,$\msun \cmq$ (shown as circles), we explore the hyperaccretion regime in this paper where the Bondi radius is larger than the \str~radius. The ionizing radiation is not able to regulate the gas accretion and becomes confined within the trapping radius \citep{Inayoshi:2016, Sakurai:2016}. The accretion is similar to the classical Bondi as the radiation from the BH is trapped in the accretion flow. Note that we normalize the gas temperature to $T_\infty = 8\times 10^3\,$K in Equations (\ref{eq:bondiradius}) and (\ref{eq:Rs}) since we allow gas cooling below $T_\infty=10^4\,$K. Even with $T_\infty=10^4$\,K as the initial condition, the temperature quickly drops to its equilibrium at $T_\infty=8 \times 10^3$\,K due to the efficient cooling from high-density gas. As a result, our selection of $\mbh$ and $\nH$ migrate to the hyperaccretion regime since $r_{\rm B} \propto T_\infty^{-1}$ and $\langle R_s \rangle \propto T_\infty^{-1/2}$. 

Table~\ref{table:para} lists the main simulation parameters. Runs in the \khp{feedback-limited} regime (M4N3 and M4N4) from PWB17 are shown for comparison. We fix the BH mass to $\mbh=10^4\,\msun$ throughout and
only change the gas density, $\nH=10^3, 10^4, 10^5$ and $10^6\,\cmq$. Integers after `N' in the simulation IDs indicate the power of gas number density as $\nH = 10^{N}\,\cmq$. Note that the simulations with the same value of
$\mbh \nH$ return qualitatively consistent results for other BH masses. For the simulations in hyperaccretion regime ($\mbh \nH \ge 10^9$\,$\msun \cmq$), we reduce the size of the simulation box $D_{\rm box}$ approximately by 
an order of magnitude as the size of the \str~sphere becomes smaller relative to M4N3 and M4N4. The box sizes are however still large enough to resolve the ionized region in the early phase of the simulations. We use a resolution of $32^3$ on the top grid with three levels of refinement for most runs, and four levels of refinement for high resolution runs which are noted as `hi' in the IDs. We attain the finest resolution of $\Delta D_{\rm min}=0.0156$\,pc for most cases ($0.0078$\,pc for high resolution runs). We force the maximum level of refinement for the volume of $(0.08\,{\rm pc})^3$ at the center of the domain. We use the gradients of gas density and gravitational potential to flag cells for refinement and outflow boundary conditions are applied on all outer boundaries.

Note that the family of M4N5 runs in Table~\ref{table:para} is located at the transition of the two accretion regimes. In feedback-limited regime, the size of the \str~sphere during the bursts of accretion is much larger than the Bondi radius as the luminosity is close to the Eddington limit. However, in the hyperaccretion regime at or above the Eddington luminosity \citep[e.g.,][]{Inayoshi:2016,Sakurai:2016}, the size of the ionized region becomes smaller with increasing gas density when the accretion luminosity is capped. \citet{Jiang:2019} show that the radiative efficiency drops to $\sim$1\% for $\dot{M}_{\rm BH}=1500\,L_{\rm Edd}/c^2$ which suggests an insensitive luminosity with an increasing accretion rate in the hyperaccretion regime.
Also we note that the size of the ionized region is not sensitive to the luminosity limit as $R_s \propto L_{\rm BH}^{1/3}$, and thus we adopt the Eddington luminosity as the limit in this paper (thick dotted-dashed line in Fig.~\ref{fig:mn}). \khp{In the Appendix~\ref{sec:appendix}, we discuss the effect of higher maximum luminosity and accretion rate on the accretion flow structure.} Fig.~\ref{fig:mn}, other possible choices of luminosity limit $10\,L_{\rm Edd}$ and $30\,L_{\rm Edd}$ are shown for a comparison (thin dotted-dashed lines).

In Fig.~\ref{fig:vr_comp}, the left two panels show radial velocities for M4N3 and M4N4 in the \khp{feedback-limited} regime where the hot ionized gas sphere regulates gas accretion. In the hyperaccretion regime such as M4N5 and M4N6 shown on the right two panels, the radiative feedback is not capable of regulating gas accretion, and thus the ionized region shrinks and radial inflow dominates. Note that the Bondi radius ($r_{\rm B} \simeq 0.65$\,pc) remains the same for all cases, however, the ionized region becomes smaller as a function of gas density, and finally leading the system to hyperaccretion regime when the density reaches a critical value.

\begin{figure}
   \includegraphics[width=\linewidth]{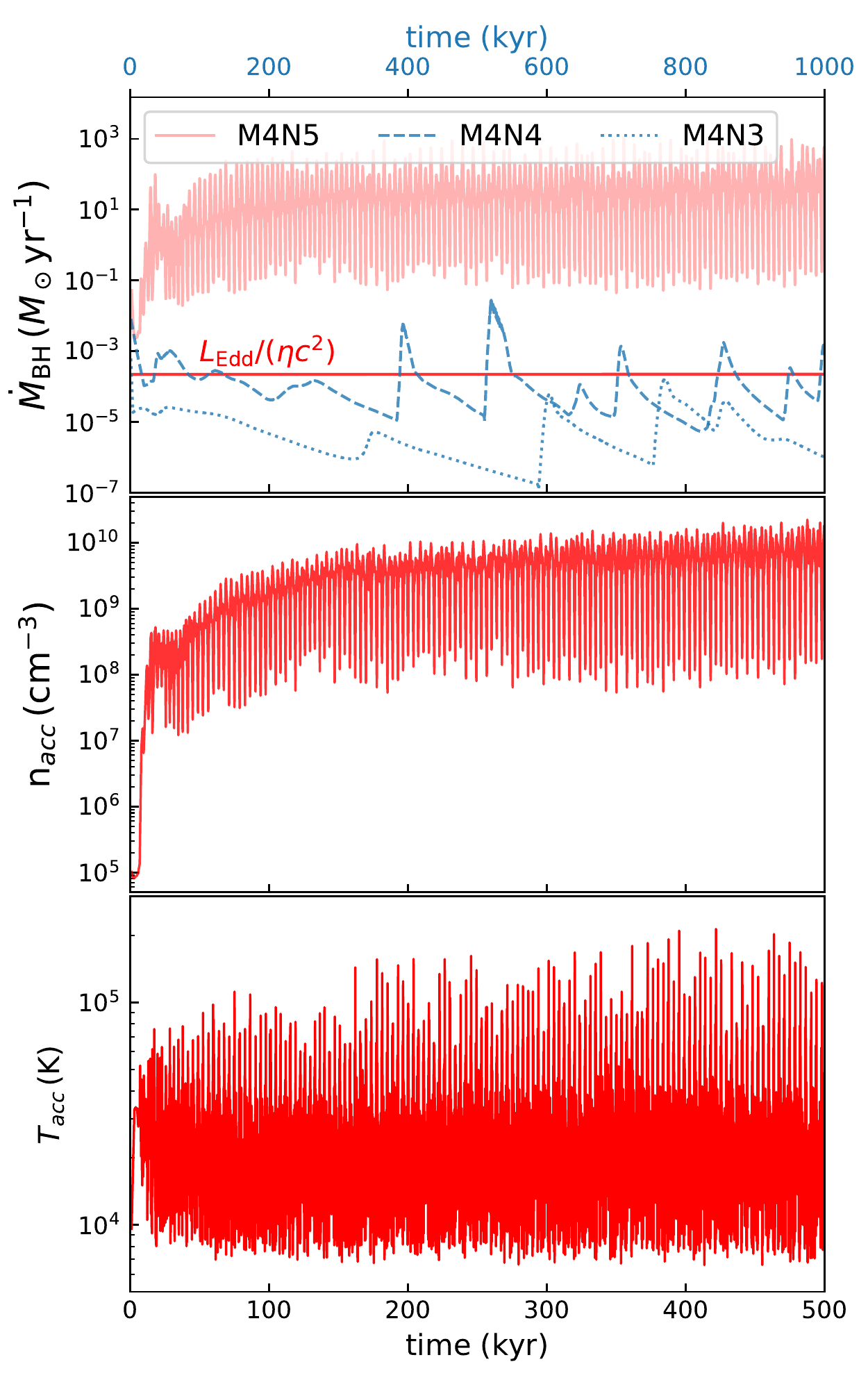}    
   \caption{Accretion rate (top) and gas density/temperature (middle/bottom) used for calculation of accretion in the run M4N5 (red). For a comparison, the M4N3 and M4N4 runs (top x-axis is used to incorporate different timescales) from PWB17 in the feedback-limited regime are shown (blue). As the $\dot{M}_{\rm BH}$ of M4N5 using Eq.(\ref{eq:acc}) is unphysically high, we apply $L_{\rm Edd}/(\eta c^2)$ as the upper limit for accretion rate.}
   \label{fig:accrate} 
\end{figure}

\begin{figure*}
   \includegraphics[width=\linewidth]{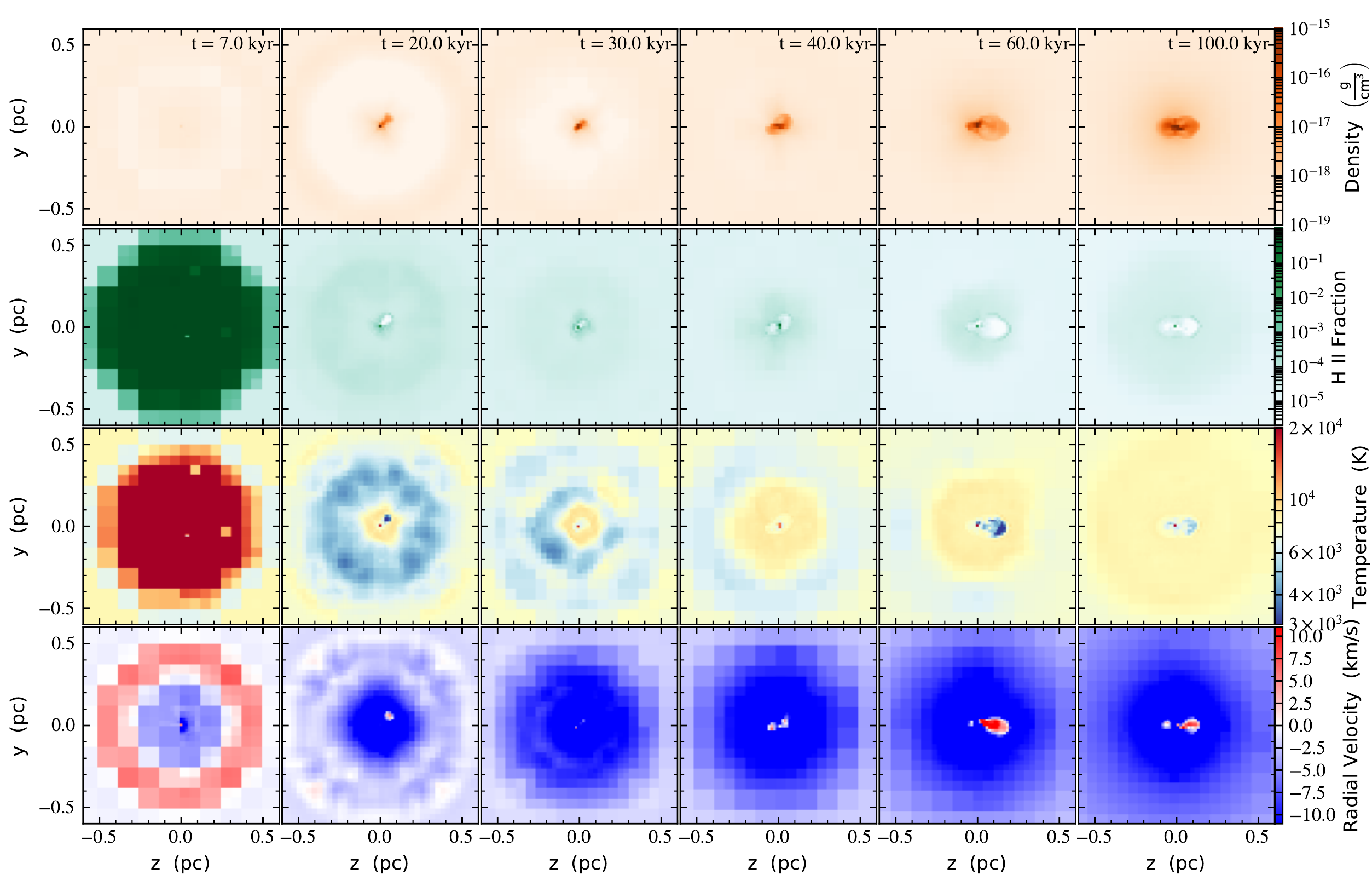} 
   \caption{Slices of M4N5c-free run at t=7, 20, 30, 40, 60 and 100\,kyr from left to
   right. From top to bottom, panels show density, \hii~fraction,
   temperature, and radial velocity snapshots. Initial formation
   of \hii~region at t=7\,kyr disappears quickly not being seen at t=20\,kyr and 
   a stochastic interaction between the BH and dense gas clumps initiates
   a rotationally supported dense structure.} 
   \label{fig:early}
\end{figure*}

\subsection{Spherical symmetry and BH position}
We explore how the spherical symmetry is affected by the BH position in the hyperaccretion flow. As mentioned earlier, in the feedback-limited regime the size of the ionized region is much larger than the Bondi radius and the spherical symmetry is maintained throughout the simulations (see PWB17). First, we test whether the symmetry is well maintained until the end of the run under an ideal condition devised for spherically symmetric accretion in the run M4N5c in Table~\ref{table:para}. We fix the BH position at (0.0078, 0.0078, and 0.0078) pc, which is the center in one of the cells adjacent to the simulation box center in the highest refinement level. After forming an ionized region quickly, the accretion structure reaches a steady state close to the Bondi accretion flow. We find that this is the only case that the spherical symmetry is maintained in the presented simulations (see the 5th column in Table~\ref{table:para}). All runs with `c' in the IDs of Table~\ref{table:para} indicate that the BH is initially centered in one of the finest cells.

We next relax the spherical symmetry of the gas inflow relative to the BH using the following methods. We simulate cases when BHs are not fixed in position. Note that the simulations with `free' in IDs such as M4N5-free and M4N6-free-hi are the runs which allow the BHs to exchange momentum with the surrounding gas. When a BH is located off the center of a spherically symmetric converging flow, the net angular momentum of the gas relative to the BH becomes nonzero. We find that this setup leads to deviations from the spherical symmetry, even with BHs initially centered in a cell as explained above. In another setup, we fix the BH at (0.004, 0.004, and 0.004)\,pc, which is off-centered on a grid and introduces a small numerical deviation from the symmetry since the gravitational potential of the BH is not symmetrically aligned relative to the grid. Simulations such as M4N5 and M4N6 in Table~\ref{table:para} are performed using this configuration.


\section{Results}
\label{sec:results}

We present our 3D simulations in the hyperaccretion regime that evolve to a stable structure that comprises a BDAF, outflowing decretion disk, and gas recycled through a meridional structure.

\subsection{Accretion rate}

The top panel of Fig.~\ref{fig:accrate} shows the accretion rate evolution while the bottom two panels show the gas number density $n_{\rm acc}$ (middle) and temperature $T_{\rm acc}$ (bottom) used to calculate the accretion rate in the M4N5 run. Here $n_{\rm acc}$ and $T_{\rm acc}$ correspond to the $\rho_{\rm HII}$ and $c_{\rm HII}$ used in Eq.~(\ref{eq:acc}) to calculate the accretion rate $\dot{M}_{\rm BH}$. At all times, this accretion rate $\dot{M}_{\rm BH}$ exceeds the Eddington rate $L_{\rm Edd}/(\eta c^2)$ (red horizontal line in the top panel of Fig.~\ref{fig:accrate}), and we use the latter when calculating the luminosity in the Eddington-limited prescription. The evolution of the accretion rate is distinct from M4N4 (blue dashed line) and M4N3 (blue dotted line) in the feedback-limited regime (PWB17), where the ionized region forms and constantly oscillates because of its varying accretion rate. In the case of M4N5, the high accretion rate is simply the result of an enhanced gas density, shown by the rapid increase in the early phase from $n_{\rm acc}=10^5\,{\rm cm}^{-3}$ to $\gtrsim 10^8\,{\rm cm}^{-3}$ in the middle panel of Fig.~\ref{fig:accrate}. The temperature of the accreted gas $T_{\rm acc}$ shows rapid fluctuations between $10^4$ and $10^5\,$K (bottom panel), but mostly cold gas with $\sim 10^4$\,K is accreted to the BH. In comparison, the temperature of the accreted gas remains high at $T \gtrsim 4\times 10^4$\,K most of the time in the cases of M4N3 and M4N4 as shown in PWB17.

\subsection{Initial \str~sphere formation}
Our simulations show that the ionizing photons create an ionized region briefly only at the beginning of all the simulations. The size of the ionized region is comparable to or smaller than the Bondi radius, which satisfies the  condition for the hyperaccretion \citep{ParkRDR:14a}. The ionized region then shrinks in size as the density increases in the central region, where the ionizing photons become trapped. The intense gas inflow toward the BH drives the system to settle down to the hyperaccretion regime, where its radiation is not able to regulate gas accretion. This runaway process continues because the BH luminosity is Eddington limited.

Fig.~\ref{fig:early} shows snapshots for the early evolution ($t \le 100\,$kyr) of density, \hii~fraction, temperature, and radial velocity from top to bottom for the M4N5c-free run. At $t=7\,$kyr (cf. the sound-crossing timescale $t_{\rm sound}=r_{\rm B}/c_\infty \approx 73$\,kyr), a spherically symmetric ionized region forms and the \str~radius extends up to $\sim 0.6\,$pc which is comparable to the Bondi radius [Eq.~(\ref{eq:bondiradius})]. The temperature of the ionized region also increases, being consistent with the \hii~fraction. The gas located in the outer part of the \str~sphere shows an outward motion (shown as red) while the gas near the BH shows inward motion (shown as blue). At this initial phase, the thermodynamic structure of the ionized region is similar to the cases of M4N3 and M4N4. However, since the subsequent accretion luminosity is not high enough to support the  \str~sphere, the ionized region shrinks in size and the gas density near the BH increases rapidly.

The second column of Fig.~\ref{fig:early} ($t=20$\,kyr) shows the snapshot after the I-front collapse. The \hii~fraction clearly shows that the ionized region retreats to the central region. The gas motion is dominated by inflow. \khp{When the BH is in a hyperaccretion state, most of the radiation cannot propagate farther than the trapping radius \citep[see][for details]{Inayoshi:2016} that is $\simeq 5\times 10^{-7}$\,pc for $\mbh=10^4$\,$\msun$, given $\dot{m}=\dot{M}_{\rm BH}/(L_{\rm Edd}/c^2)=10^3$. At this level, the trapping radius is over five orders of magnitude below our resolution limit. Thus, the vast majority of the ionizing radiation remains confined within the computational cell that hosts the BH.} However, some radiation preferentially leaks into low density direction since the density is not homogeneous near the BH. The direction and mean free path of the escaping radiation are highly variable, creating a partially ionized central region, significantly smaller than the initial \str~radius. 


\begin{figure*}
   \begin{center}
   \includegraphics[width=\linewidth]{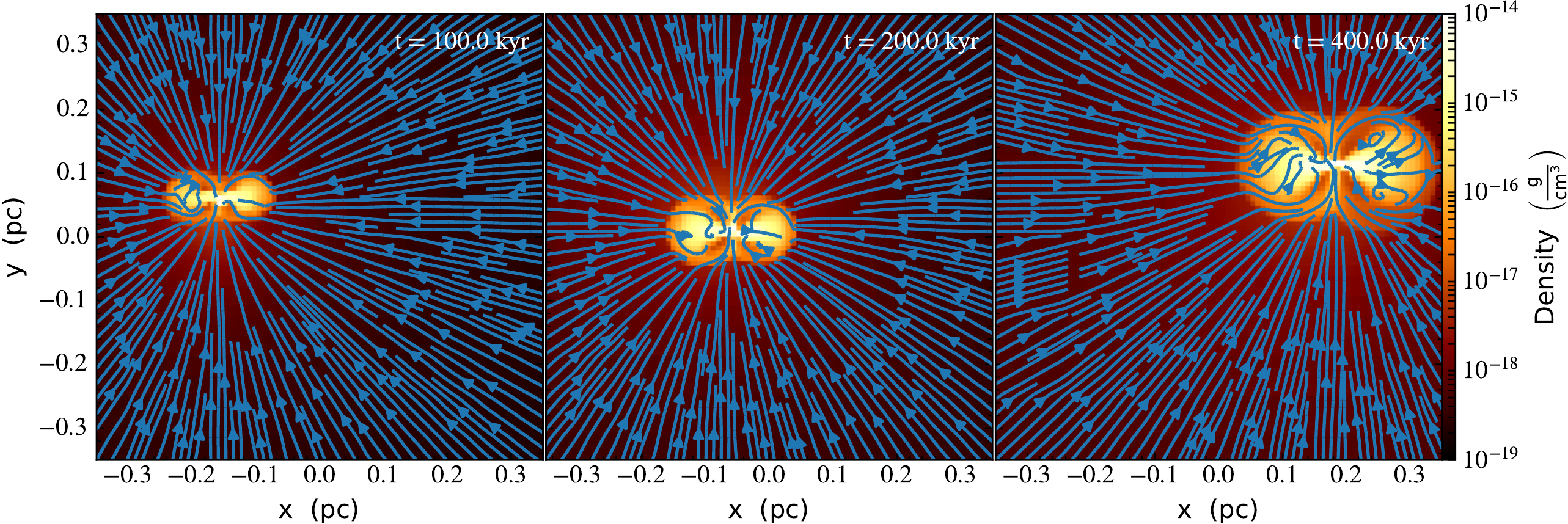}
   \caption{Time sequence of density slices at $z=z_{\rm BH}$ where $z_{\rm BH}$ is the z-coordinate of the BH plotted together with velocity streams of the accretion flow for M4N5c-free-hi run. Radial inflow dominates in the early phase of the simulation while BDAF develops near the BH and the structure becomes reinforced progressively. Note that (0, 0) is the domain center of x-y plane and the location of the BH is not fixed. }
   \label{fig:velo_xy}
\end{center}
\end{figure*}

After a period of stochastic interactions between the radiation and high-density gas, a rotating component perpendicular to the y-axis starts to form and increases in size. The second column of Fig.~\ref{fig:early} shows a dense gas clump, which passes by the BH and moves away from it. This dense and cold gas clump is bound by the gravity of the BH and continues to orbit around it. The radial velocity slice at $t= 60\,$kyr clearly shows an enhanced outflow structure, which is rotationally supported. The overall gas motion shows an inward velocity (blue) toward the BH, similar to Bondi accretion. On the other hand, an outflow (red) perpendicular to the y-axis starts to develop in the vicinity of the BH. We find that the orientation of rotational axis is random and varies among the simulations listed in Table.~\ref{table:para}.

\begin{figure*}
   \begin{center}
   \includegraphics[width=0.6\linewidth]{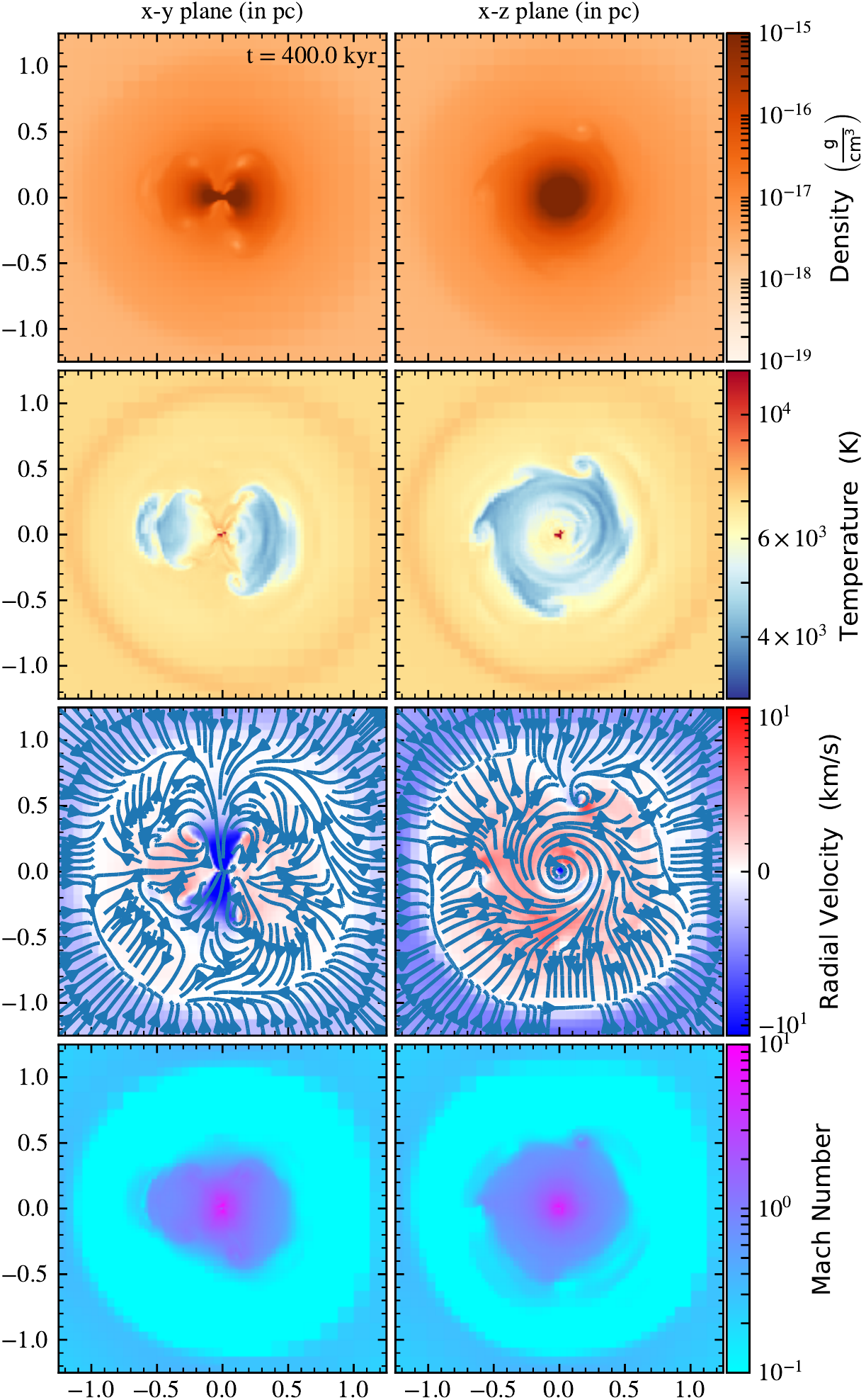}
   \caption{Slices of density, gas temperature, radial velocity, and Mach number of
   BDAF and {\it decretion} disk of M4N6 run from top to bottom: left panels 
   show the structure of the BDAF along $y$-axis in the $x$-$y$ plane whereas right panels 
   show the {\it decretion} disk in the $x$-$z$ plane.}
   \label{fig:xyz}
   \end{center}
\end{figure*}

\begin{figure*} 
\plottwo{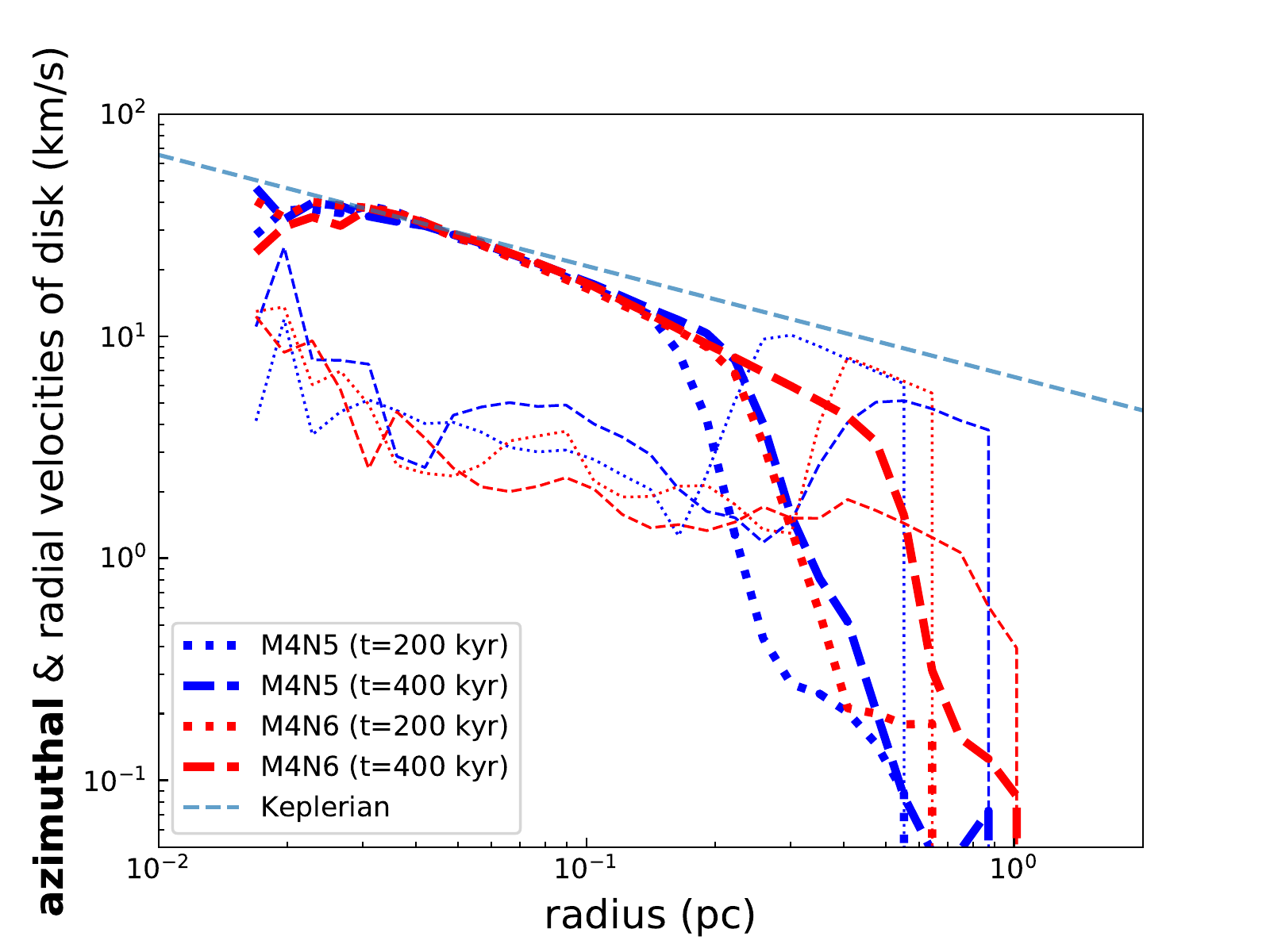}{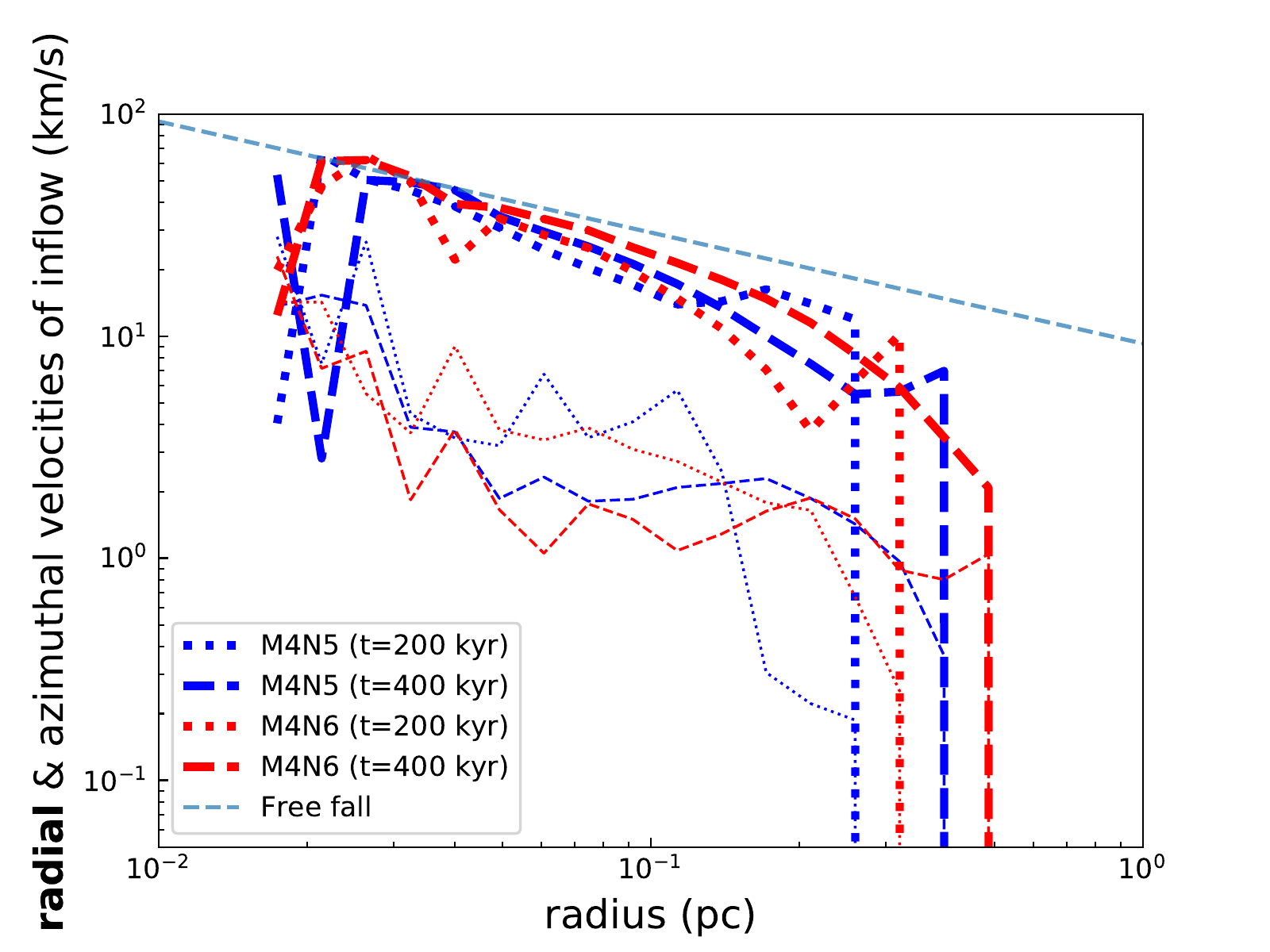}
\caption{Left: absolute mass-weighted average azimuthal (thick) and radial (thin) velocity profiles of rotating disk for M4N5 (blue) and M4N6 (red) runs at different epochs of t=200\,kyr (dotted) and 400\,kyr (dashed). Right: radial (thick) and azimuthal (thin) velocity profiles of biconical inflow for the same runs. }
\label{fig:velo_profile}
\end{figure*}

\subsection{Growth of a biconical inflow and decretion disk}
Fig.~\ref{fig:velo_xy} shows the evolution of velocity structure in the $x$-$y$ plane for M4N5c-free-hi run from t=100 to 400\,kyr. The figure shows the edge-on view of the density structure plotted together with velocity streamlines. This time sequence shows that the structure grows in size and drifts to the right due to the momentum exchange with \khp{the} surrounding gas. At $t=100$\,kyr, the inflow onto the central BH dominates and \khp{an initially weak outflow starts to develop in the rotating disk.} As the outflow along the rotating disk becomes stronger, \khp{it} starts to interfere with the inflow. However, the gas inflow along the y-axis still freely falls into the central region developing a biconical gas inflow as shown at $t=200$\,kyr. The biconical inflow and rotationally supported dense disk grow in size, and two flows interact with each other only at the interface. The mass of the rotating disk increases and the solid angle of the funnels containing inflowing gas decreases. A large fraction of mass from the biconical inflow is deflected to the bulk rotational outflow. The two distinct regions are characterized by opposite directions of radial velocity shown in Fig.~\ref{fig:velo_xy}. At $t=400\,$kyr, the structure of the flow reaches a steady state and the biconical inflow still exists.

\subsection{Hydrodynamic structure of biconical inflow and decretion disk}

Fig.~\ref{fig:xyz} shows the 3D structure of density, temperature, radial velocity, and Mach number at $t=400\,$kyr for M4N6 run. The left panels are slices in the $x$-$y$ plane while the right panels are in the $x$-$z$ plane. The overall structures are similar to the case of M4N5c-free-hi shown in Fig.~\ref{fig:velo_xy}, however they show some features that are distinct due to the higher density. A dense rotational disk and low-density channel of BDAF is seen in the top left panel, which is clearly characterized in the radial velocity slices. 

At large scales, the rotational outflow forms a quasi-spherical interface with the inflow at $r \sim 0.5$\,pc. The inflows from large scales are stopped by the outflow, but the biconical structure remains as a meridional structure develops. Some of the outflow becomes recycled for the BH accretion by being deflected into the biconical channel. The structure of the rotational disk and biconical inflow is not perfectly axisymmetric; however, this feature may arise from accumulating numerical errors. The rotational disk has an elevated density, enhancing its cooling and resulting in a lower temperature distribution inside it shown in the temperature slices. The outer edge of the disk shows the characteristics of a Kelvin-Helmholtz instability caused by the shear between the inflow and outflow. 



The collision of the BDAFs creates density waves that propagate outward from the central region. The slices of Mach number for total velocity show that the central region is highly supersonic and the edge of the rotational disk at $\sim$\,0.5\,pc shows a transonic feature. The high-resolution run with the BH not fixed (M4N6-free-hi) also shares qualitatively same results, however, with a more compact decretion disk.

Fig.~\ref{fig:velo_profile} shows the velocity profiles of the {\it decretion} disk (left) and BDAF (right). The left panel shows the the mass-weighted average azimuthal (thick) and radial (thin) velocity profiles of a cylinder with a height of $0.0156$\,pc centered at the BH for M4N5 (blue) and M4N6 (red) runs. We select two epochs at $t=200$\,kyr (dotted) and $400$\,kyr (dashed) to show the growth of the structures. The inner region at $r < 0.1$\,pc displays a similar profile to Keplerian motion due to the gravity of the BH. The profiles gradually deviate from \khp{Keplerian} at larger radii. The azimuthal velocities drop rapidly below $v_\phi = 1.0\,{\rm km\,s^{-1}}$ at $t=200$\,kyr at $r \sim 0.2$\,pc where the outflows meet inflows. This transition from nearly Keplerian occurs at smaller radius for M4N5 compared to M4N6. The transition radius for M4N6 run continues to increase to $r \sim 0.5$\,pc at $t=400$\,kyr. The difference in velocity transition is visually consistent with the one in hydrodynamic structures observed in Fig.~\ref{fig:velo_xy} for M4N5 and Fig.~\ref{fig:xyz} for M4N6. The radial velocity profiles are approximately one order of magnitude lower than azimuthal velocities.

Similarly, radial (thick) and azimuthal (thin) velocity profiles of the gas along a \khp{line} parallel to the biconical inflow centered at the BH are shown in the right panel of Fig.~\ref{fig:velo_profile}. The radial velocity of the biconical inflow follows the freefall velocity which continues up to the resolution limit where gas from the opposite directions of the biconical channel collide. The accretion flow falling through the biconical channel interacts with the gas on the surface of the rotating disk. As a result, Kelvin-Helmholtz instabilities also develop on the interface of the two regions. The radial inflow velocities become affected by the gas, which turns around from the decretion disk. This gas does not originate form the outer part of the simulation box, but is just recycled from the disk. In the biconical inflow, the radial velocity dominates over the azimuthal component in contrast to the case of decretion disk.


\subsection{Net angular momentum}
Our simulations start from the zero angular momentum initial condition; however, the net angular momentum evolves since the gas is accelerated under the influence of the BH gravity. After the collision of the two gas inflows, gas with nonzero angular momentum relative to the BH is redirected to the rotating disk. The net angular momentum is deposited in the disk and increases as a function of time as the disk builds up in mass. This does not occur for M4N5c where the spherical symmetry is maintained throughout the run. In other runs, we find that \khp{a} deviation from the spherical symmetry results in increased net angular momentum.

\begin{figure} 
   \includegraphics[width=\linewidth]{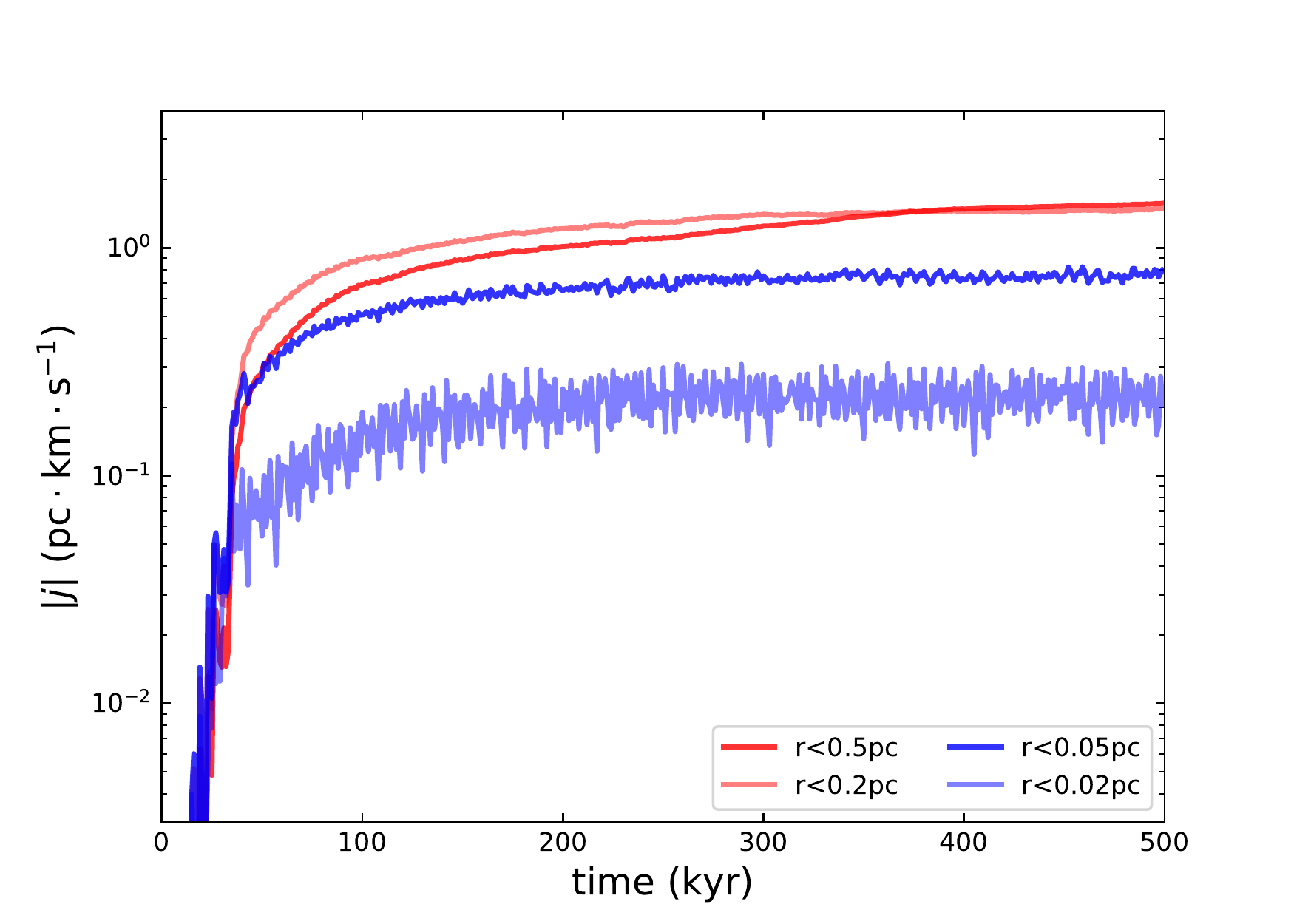}
   \caption{Evolution of specific angular momentum of gas \khp{for M4N6 run}: red lines for
   the angular momenta for gas within the large radii ($r \ge 0.2$\,pc)
   and blue lines show for small radii ($r \le 0.05$\,pc).}
   \label{fig:L_evol}
\end{figure}

Fig.~\ref{fig:L_evol} shows the evolution of specific angular momentum of gas $|j|$ \khp{for M4N6 run}. The net angular momentum of the box increases monotonically until $t \sim 300$\,kyr (red lines). The specific angular momentum within \khp{the spheres with} small radii (blue lines within $r < 0.05$\,pc) becomes saturated at earlier times ($t \sim 200$\,kyr) and does not increase afterward. Only gas with a vanishingly small angular momentum can approach close to the BH, and gas with greater angular momentum is redirected to the rotating disk. Thus, regardless of its source, the angular momentum is stored in the rotating disk, whose mass stops increasing at $t \sim
300$\,kyr. At this point of transition, the size of the decretion disk becomes similar to the Bondi radius and stops increasing. The asymptotic value of $|j|$ for $r<0.5$\,pc is consistent with Fig.~\ref{fig:velo_profile}, as $j=v_\phi r \sim (10\,{\rm km}\,{\rm s^{-1}})$($0.1$\,pc) $\sim 1\,{\rm pc}\cdot {\rm km}\cdot {\rm s}^{-1}$.

\begin{figure}
   \includegraphics[width=\linewidth]{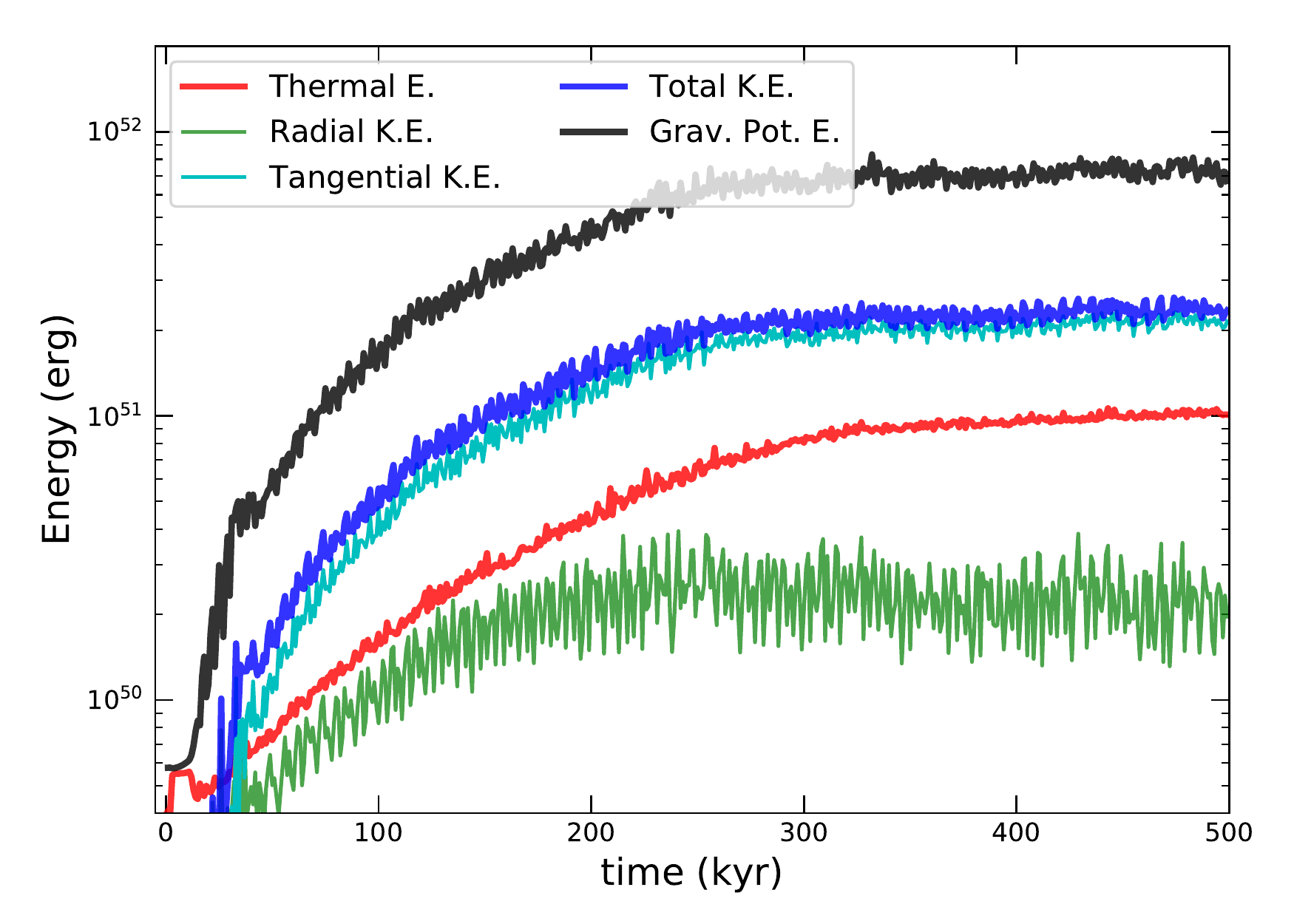}
   \caption{Evolution of kinetic, thermal, and gravitational potential energy of gas within $r_{\rm B} = 0.65$\,pc as a function of time for M4N6 run.
   }
   \label{fig:energy_evol}
\end{figure}

Fig.~\ref{fig:energy_evol} shows the evolution of kinetic, thermal, and gravitational potential energy of the gas within the Bondi radius $r_{\rm B}$ of M4N6 run. Kinetic energy (blue) increases monotonically until $t=300$\,kyr and stays at $E_{\rm kin} \sim 2 \times 10^{51}$\,erg. The thermal energy of the gas (red) starts from a higher value; however, it increases slowly as the system equilibrates. Thus, the total K.E. starts to dominate the thermal energy. 
In contrast, in \khp{feedback-limited} regime, PWB17 show that the total energy of the system is dominated by the thermal energy of the gas. The total kinetic energy, including turbulence was only a small fraction of the total thermal energy in that case. However, in the hyperaccretion regime we find that the system reaches a steady state that is dominated by the gas motion. The radial component of K.E. (green) is approximately one order of magnitude smaller than the tangential component of K.E. (cyan). The simulation shows that the total kinetic energy \khp{is} about twice of the thermal energy at the end of the run at $t=500$\,kyr.

\section{\khp{Discussion}}
\label{sec:discussion}

The setup of the current simulations is still idealized in several respects and the presented results should be carefully understood and tested in a more realistic setup in cosmological simulations in the future. For example, we do not update the BH mass as only a few percent of mass would be increased with the Eddington-limited growth rate within the simulation time (i.e., 500 kyr). 

\khp{Higher accretion rates might be possible as long as the accretion rate is consistent with the geometry of BDAF.} \khpii{The fraction of the solid angle of BDAF multiplied by the Bondi accretion rate can lead to $\gtrsim$\,10 times higher accretion rate for the M4N6 run, compared to the M4N5 run,} which is potentially important for the rapid growth scenario of the seed BHs in the early universe (see Appendix~\ref{sec:appendix}).
\khp{We use the Eddington luminosity as the cap; however, a broader exploration is necessary in the future to study the transition criterion between feedback-limited and feeding-dominated regimes shown in Fig.~\ref{fig:mn}. Our preliminary test with a higher luminosity cap of $10\,L_{\rm Edd}$ for M4N5 run, motivated by \citet{Jiang:2014, Jiang:2019}, creates an approximately 2 times larger \str~sphere in the beginning, which is greater than the Bondi radius as $\langle R_s \rangle \propto L_{\rm BH}^{1/3}$. However, the collapse of the \str~sphere still occurs as the outflow inside the \str~sphere is not strong enough to reduce the gas density within the \str~sphere as the cases in the \khp{feedback-limited} regime (see the Appendix~\ref{sec:appendix}).} 
We also do not include the effect of self-gravity of the gas. Because the total mass of the rotating disk is significantly larger than the BH mass at the end of the simulations, meaning that the disk might be unstable to gravitational collapse. If so, local gravitational instability in the disk may lead to star formation and global instabilities (like spiral arms) may provide a way to transport angular momentum. Neither of these effects is captured in our study.

The current study does not explore in quantitative detail how angular momentum is generated and transferred.
Nonetheless, we describe it in a qualitative way. The local angular momentum of gas inflow relative to the BH can deviate from zero when the BH is off-centered from the convergence point that is moving with a certain velocity. This process occurs due to the time delay for the gas inflow from large scale to reach the BH. As a result, when the gas with nonzero angular momentum reaches the BH, it misses the BH and is deflected to the rotational disk. With a rapid increase in density, the disk cools and becomes stable. This process stores the gas mass and thus angular momentum carried by the gas into the disk. We do not model the angular momentum transfer mechanism due to viscosity or magnetorotational instability, so any angular momentum transport in our simulations is a consequence of gravitational and fluid interactions.

For these reasons, we limit the scope of the current work to the geometry of the flow and demonstrate that a spherical geometry is unstable in the hyperaccretion regime. We extensively test whether the BDAF \khp{geometry} is \khp{a} purely numerical \khp{outcome} when we fix the BH position. We confirm that when the BH is allowed to move freely by exchanging momentum with the surrounding gas, the BH drifts away from the center of the simulation box at a constant velocity due to the momentum acquired from interactions with high-density gas at early times. A stable axisymmetric structure forms during this phase and survives until the end of the simulation. 

The innermost structure of the biconical inflow might be an interesting topic for the future work. With higher resolution simulations, we can test how far the structure extends beyond the current resolution limit. For example, in order to resolve $100\,r_g \sim 3\times 10^6\,{\rm km}$ where $r_g$ is the gravitational radius for a BH mass $\mbh=10^4$\,$\msun$, which can provide the boundary condition for magnetohydrodynamic simulations of accretion disks, we need to increase the resolution by $\sim 5$ orders of magnitude. 

The phenomenon described in this work might be applicable to other astrophysical scenarios. In theory, any gravitating object accreting gas at a high rate with nonzero angular momentum might expect the same result. For example, a similarity exists in a theory for a newly formed giant planet embedded in a circumstellar disk. \citet{Batygin:2018} explains the low terminal rotation of Jupiter using a meridional circulation of gas that connects the decretion disk and bipolar accretion flow within the Hill sphere of the planet. 

\section{\khp{Summary}}
\label{sec:summary}
We perform 3D radiation-hydrodynamic simulations using {\it Enzo} equipped with {\it Moray} to study BHs in the hyperaccretion regime. We put an emphasis on a stable hydrodynamic structure that deviates from spherical symmetry. We list our main discoveries in the following points. 

\begin{itemize}  
   \item Our simulations show that ionizing radiation from a BH in the hyperaccretion regime is not able to regulate gas accretion since the ionizing photons are trapped within the accretion inflow. \khp{The ionized region which forms initially is comparable to or smaller than the Bondi radius}, but then shrinks to the resolution limit.
   
   \item  The steady state hyperaccretion flow is not spherically symmetric in the general case. When we relax the assumption of spherical symmetry, random interactions between the BH and nonuniform gas lead to a formation of an axisymmetric structure triggering off a rotating gas cloud around the BH.
   
   \item Once spherical symmetry is broken, a nearly radial BDAF develops perpendicular to the rotating disk, avoiding the centrifugal barrier. The BDAFs from the opposite directions collide near the BH and only the gas with the smallest angular momentum reaches the BH, whereas the rest is deflected into the rotating disk. The {\it decretion} disk becomes progressively reinforced as more mass from the biconical flow transfers to the disk.
   
   \item As the decretion disk grows in mass, some gas on the surface of the disk is recycled into the biconical accretion channels developing a meridional gas flow. This axisymmetric structure provides a stable configuration that can deliver an uninterrupted supply of high density gas to the BH. This process continues until the size of the outflow decretion disk becomes comparable to the scale of Bondi radius.      
\end{itemize}

In conclusion, hyperaccretion provides a cogent solution for the rapid growth of seed BHs in the early universe. Our current study suggests a possibility for a stable 3D configuration of a hyperaccretion flow on the scale of the Bondi radius. Our simulations are characterized by a stable BDAF perpendicular to any rotating disk that might form out of stochastic interactions between the BH and non-uniform gas. The \khp{BDAF} might last until all the nearby gas within the Bondi radius is consumed by the BH as the meridional structure recycles gas from rotational disk to biconical channel. 

\acknowledgements
The presented work is supported by the National Science Foundation (NSF) grants AST-1614333 and OAC-1835213, and National Aeronautics and Space Administration (NASA) grants NNX17AG23G and 80NSSC20K0520. T.B. acknowledges the support by the NASA under award No. 80NSSC19K0319 and by the NSF under award No. 1908042. K.P. thanks Konstantin Batygin for private discussion. Numerical simulations presented were performed using the open-source {\it Enzo} and the visualization package {\sc yt} \citep{Turk:2011}.

\begin{figure}
   \begin{center}
   \includegraphics[width=0.8\linewidth]{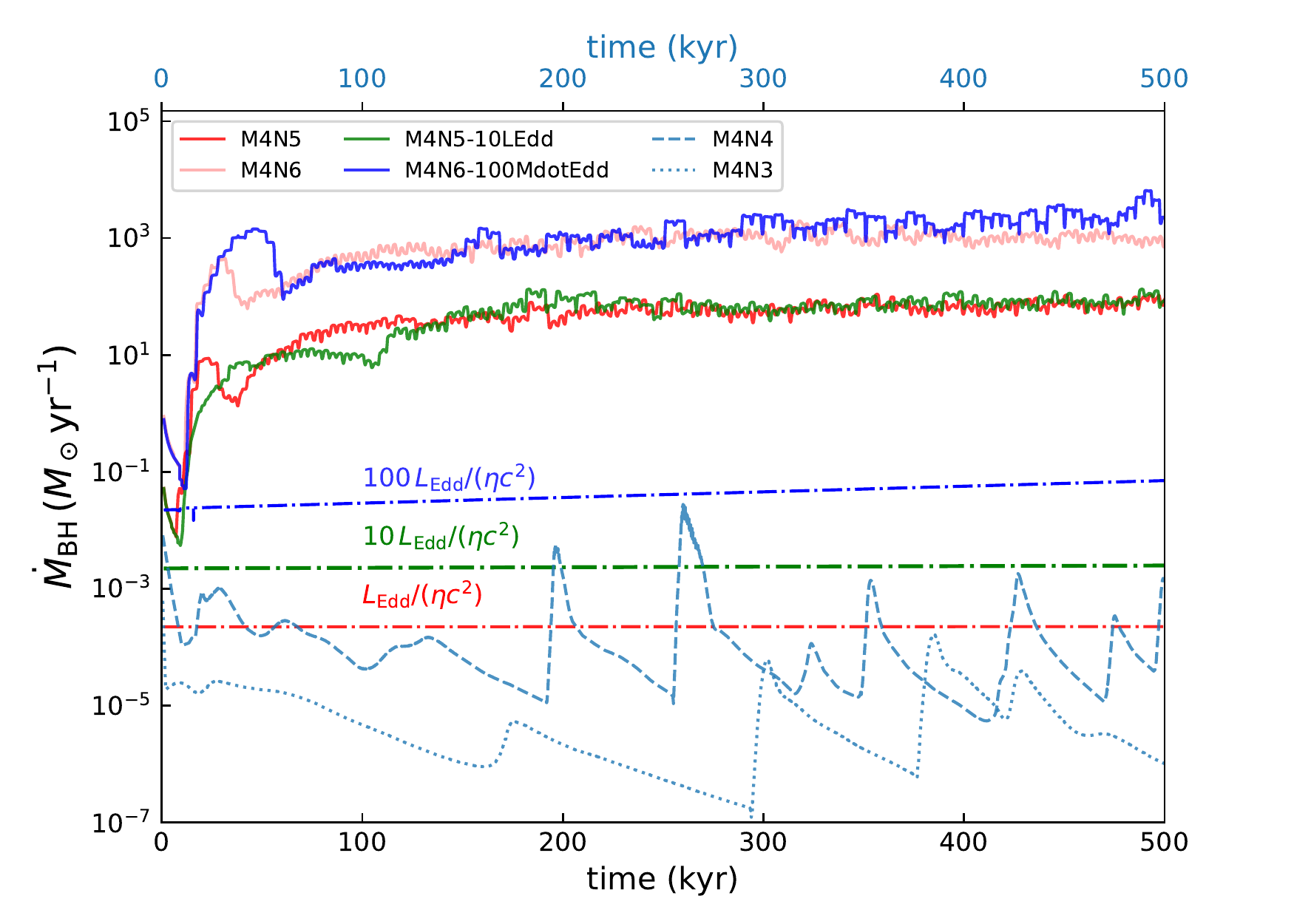}    
   \caption{Evolution of estimated Bondi accretion rates in M4N5, M4N5-10LEdd, M4N6, and M4N6-100MdotEdd with moving averages with a window of $\sim$10\,kyr (solid lines) and corresponding maximum accretion rate models $L_{\rm Edd}/(\eta c^2)$, $10L_{\rm Edd}/(\eta c^2)$ and $100L_{\rm Edd}/(\eta c^2)$ (dotted-dashed lines).}
   \label{fig:superEdd}
   \end{center} 
\end{figure}

\khp{
\appendix

\section{Super-Eddington luminosity and accretion rate}
\label{sec:appendix}

In the current study, we make a conservative assumption for the maximum luminosity as $L_{\rm max}=L_{\rm Edd}$ which makes M4N5 runs in the hyperaccretion regime as shown in Fig.~\ref{fig:mn}. Adopting $L_{\rm max}$ approximately 10 times higher motivated by MHD simulations by \citet{Jiang:2014, Jiang:2019}, might make M4N5 shift to the \khp{feedback-limited} regime. However, Fig.~\ref{fig:mn} assumes that the gas temperature is $T_{\infty}=10^4$\,K to incorporate feedback-limited regime where the gas stays at $T_{\infty}=10^4$\,K due to less cooling. Since the temperature drops quickly to the temperature floor of $T_\infty \simeq 8000$\,K due to rapid gas cooling in the case of high-density runs, the transition line shown in Fig.~\ref{fig:mn} shifts downward making the M4N5 run stay marginally inside the hyperaccretion regime. This is due to the fact that the Bondi radius is proportional to $T_{\infty}^{-1}$ and the mean size of \str~radius is $\ars \propto T_{\infty}^{-1/2}$. Therefore, the effect of $L_{\rm max}=10 L_{\rm Edd}$ is reduced due to the lower temperature $T_\infty$.

We run a simulation M4N5-10LEdd with $L_{\rm max}=10\,L_{\rm Edd}$ (shown as green lines in Fig.~\ref{fig:superEdd} which increase the initial size of the \str~radius by a factor of $10^{1/3}\sim 2.15$, which makes the initial $\ars \sim 1\,$pc greater than the Bondi radius $r_{\rm B} \sim 0.7$\,pc. Despite the increased $\ars$, we find that M4N5-10LEdd run still stays in the hyperaccretion regime. The current border line in Fig.~\ref{fig:mn} for the hyperaccretion regime is based on the condition $\ars \lesssim r_{\rm B}$. In the run M4N5-10LEdd, the initial $\ars$ is greater than $r_{\rm B}$ by approximately $1.5$ times; however, the system evolves to a hyperaccretion state as $\ars$ cannot stay in a steady state due to the weak outflow inside the \str~sphere. In high-density runs, the thermal pressure gradient inside the \str~sphere is not steep enough to produce a strong outflow, which is the key to the stability of the \str~sphere. Fig.~6 in \citet{ParkRDR:14a} also shows that $\ars \sim 2 r_{\rm B}$ which stays in a steady state initially, but eventually the size of the ionized region shrinks. 

We also apply the Eddington-limited growth rate in the current study; however, there is a possibility that the actual rate could be much larger than that. Our current simulations are limited in constraining the actual accretion rate since they do not show how the accretion flow structure changes as it gets closer to accretion disk scales. Nonetheless, for a steady-state solution, the actual accretion rate should be consistent with the BDAF structure, if it extends to smaller scales. The solid angle of the BDAF, which is a few percent of the entire solid angle, indicates that a few percent of Bondi accretion rate might be consistent with current selection of accretion rate cap for M4N5 runs. However, with increasing gas density, the corresponding Bondi rate will increase, which indicates that the accretion rate might be $\sim$\,10 times of the Eddington-limited value (e.g., for M4N6 runs). We run another simulation, named M4N6-100MdotEdd, with a 2 orders of magnitude larger cap on the accretion rate but with the same $L_{\rm max}=L_{\rm Edd}$ (shown as blue in Fig.~\ref{fig:superEdd}) and find that the BDAF is still found. The final mass of the BH at $t=500$\,kyr is about 3 times the initial mass. This opens a possibility for a rapid growth for seed BHs, but a deeper investigation of accretion flow on small scales is necessary in the future.
}

\bibliographystyle{aasjournal}
\bibliography{park_bh}

\begin{thebibliography}{}
\expandafter\ifx\csname natexlab\endcsname\relax\def\natexlab#1{#1}\fi
\providecommand{\url}[1]{\href{#1}{#1}}
\providecommand{\dodoi}[1]{doi:~\href{http://doi.org/#1}{\nolinkurl{#1}}}
\providecommand{\doeprint}[1]{\href{http://ascl.net/#1}{\nolinkurl{http://ascl.net/#1}}}
\providecommand{\doarXiv}[1]{\href{https://arxiv.org/abs/#1}{\nolinkurl{https://arxiv.org/abs/#1}}}

\bibitem[{{Abel} {et~al.}(2000){Abel}, {Bryan}, \& {Norman}}]{AbelBN:00}
{Abel}, T., {Bryan}, G.~L., \& {Norman}, M.~L. 2000, \apj, 540, 39,
  \dodoi{10.1086/309295}

\bibitem[{{Abramowicz} \& {Zurek}(1981)}]{Abramowicz:1981}
{Abramowicz}, M.~A., \& {Zurek}, W.~H. 1981, \apj, 246, 314,
  \dodoi{10.1086/158924}

\bibitem[{{Alexander} \& {Natarajan}(2014)}]{AlexanderN:2014}
{Alexander}, T., \& {Natarajan}, P. 2014, Science, 345, 1330,
  \dodoi{10.1126/science.1251053}

\bibitem[{{Aykutalp} {et~al.}(2014){Aykutalp}, {Wise}, {Spaans}, \&
  {Meijerink}}]{Aykutalp:2014}
{Aykutalp}, A., {Wise}, J.~H., {Spaans}, M., \& {Meijerink}, R. 2014, \apj,
  797, 139, \dodoi{10.1088/0004-637X/797/2/139}

\bibitem[{{Ba{\~n}ados} {et~al.}(2018){Ba{\~n}ados}, {Venemans},
  {Mazzucchelli}, {Farina}, {Walter}, {Wang}, {Decarli}, {Stern}, {Fan},
  {Davies}, {Hennawi}, {Simcoe}, {Turner}, {Rix}, {Yang}, {Kelson}, {Rudie}, \&
  {Winters}}]{Banados:2018}
{Ba{\~n}ados}, E., {Venemans}, B.~P., {Mazzucchelli}, C., {et~al.} 2018, \nat,
  553, 473, \dodoi{10.1038/nature25180}

\bibitem[{{Batygin}(2018)}]{Batygin:2018}
{Batygin}, K. 2018, \aj, 155, 178, \dodoi{10.3847/1538-3881/aab54e}

\bibitem[{{Begelman}(1979)}]{Begelman:79}
{Begelman}, M.~C. 1979, \mnras, 187, 237

\bibitem[{{Begelman}(2012)}]{Begelman:2012a}
---. 2012, \apjl, 749, L3, \dodoi{10.1088/2041-8205/749/1/L3}

\bibitem[{{Begelman} {et~al.}(2006){Begelman}, {Volonteri}, \&
  {Rees}}]{BegelmanVR:06}
{Begelman}, M.~C., {Volonteri}, M., \& {Rees}, M.~J. 2006, \mnras, 370, 289,
  \dodoi{10.1111/j.1365-2966.2006.10467.x}

\bibitem[{{Boekholt} {et~al.}(2018){Boekholt}, {Schleicher}, {Fellhauer},
  {Klessen}, {Reinoso}, {Stutz}, \& {Haemmerl{\'e}}}]{Boekholt:2018}
{Boekholt}, T.~C.~N., {Schleicher}, D.~R.~G., {Fellhauer}, M., {et~al.} 2018,
  \mnras, 476, 366, \dodoi{10.1093/mnras/sty208}

\bibitem[{{Bondi}(1952)}]{Bondi:52}
{Bondi}, H. 1952, \mnras, 112, 195

\bibitem[{{Bromm} {et~al.}(1999){Bromm}, {Coppi}, \& {Larson}}]{BrommCL:99}
{Bromm}, V., {Coppi}, P.~S., \& {Larson}, R.~B. 1999, \apjl, 527, L5,
  \dodoi{10.1086/312385}

\bibitem[{{Brummel-Smith} {et~al.}(2019){Brummel-Smith}, {Bryan}, {Butsky},
  {Corlies}, {Emerick}, {Forbes}, {Fujimoto}, {Goldbaum}, {Grete}, {Hummels},
  {Kim}, {Koh}, {Li}, {Li}, {Li}, {OShea}, {Peeples}, {Regan}, {Salem},
  {Schmidt}, {Simpson}, {Smith}, {Tumlinson}, {Turk}, {Wise}, {Abel},
  {Bordner}, {Cen}, {Collins}, {Crosby}, {Edelmann}, {Hahn}, {Harkness},
  {Harper-Clark}, {Kong}, {Kritsuk}, {Kuhlen}, {Larrue}, {Lee}, {Meece},
  {Norman}, {Oishi}, {Paschos}, {Peruta}, {Razoumov}, {Reynolds}, {Silvia},
  {Skillman}, {Skory}, {So}, {Tasker}, {Wagner}, {Wang}, {Xu}, \&
  {Zhao}}]{Brummel-Smith:2019}
{Brummel-Smith}, C., {Bryan}, G., {Butsky}, I., {et~al.} 2019, The Journal of
  Open Source Software, 4, 1636, \dodoi{10.21105/joss.01636}

\bibitem[{{Bryan} {et~al.}(2014){Bryan}, {Norman}, {O'Shea}, {Abel}, {Wise},
  {Turk}, {Reynolds}, {Collins}, {Wang}, {Skillman}, {Smith}, {Harkness},
  {Bordner}, {Kim}, {Kuhlen}, {Xu}, {Goldbaum}, {Hummels}, {Kritsuk}, {Tasker},
  {Skory}, {Simpson}, {Hahn}, {Oishi}, {So}, {Zhao}, {Cen}, {Li}, \& {Enzo
  Collaboration}}]{Bryan:2014}
{Bryan}, G.~L., {Norman}, M.~L., {O'Shea}, B.~W., {et~al.} 2014, \apjs, 211,
  19, \dodoi{10.1088/0067-0049/211/2/19}

\bibitem[{{Choi} {et~al.}(2013){Choi}, {Shlosman}, \& {Begelman}}]{ChoiSB:13}
{Choi}, J.-H., {Shlosman}, I., \& {Begelman}, M.~C. 2013, \apj, 774, 149,
  \dodoi{10.1088/0004-637X/774/2/149}

\bibitem[{{Davies} {et~al.}(2011){Davies}, {Miller}, \&
  {Bellovary}}]{Davies:2011}
{Davies}, M.~B., {Miller}, M.~C., \& {Bellovary}, J.~M. 2011, \apjl, 740, L42,
  \dodoi{10.1088/2041-8205/740/2/L42}

\bibitem[{{Devecchi} \& {Volonteri}(2009)}]{Devecchi:2009}
{Devecchi}, B., \& {Volonteri}, M. 2009, \apj, 694, 302,
  \dodoi{10.1088/0004-637X/694/1/302}

\bibitem[{{Fan} {et~al.}(2001){Fan}, {Narayanan}, {Lupton}, {Strauss}, {Knapp},
  {Becker}, {White}, {Pentericci}, {Leggett}, {Haiman}, {Gunn}, {Ivezi{\'c}},
  {Schneider}, {Anderson}, {Brinkmann}, {Bahcall}, {Connolly}, {Csabai}, {Doi},
  {Fukugita}, {Geballe}, {Grebel}, {Harbeck}, {Hennessy}, {Lamb}, {Miknaitis},
  {Munn}, {Nichol}, {Okamura}, {Pier}, {Prada}, {Richards}, {Szalay}, \&
  {York}}]{Fan:2001}
{Fan}, X., {Narayanan}, V.~K., {Lupton}, R.~H., {et~al.} 2001, \aj, 122, 2833,
  \dodoi{10.1086/324111}

\bibitem[{{Inayoshi} {et~al.}(2016){Inayoshi}, {Haiman}, \&
  {Ostriker}}]{Inayoshi:2016}
{Inayoshi}, K., {Haiman}, Z., \& {Ostriker}, J.~P. 2016, \mnras, 459, 3738,
  \dodoi{10.1093/mnras/stw836}

\bibitem[{{Inayoshi} {et~al.}(2015){Inayoshi}, {Visbal}, \&
  {Kashiyama}}]{Inayoshi:2015a}
{Inayoshi}, K., {Visbal}, E., \& {Kashiyama}, K. 2015, \mnras, 453, 1692,
  \dodoi{10.1093/mnras/stv1654}

\bibitem[{{Jiang} {et~al.}(2014){Jiang}, {Stone}, \& {Davis}}]{Jiang:2014}
{Jiang}, Y.-F., {Stone}, J.~M., \& {Davis}, S.~W. 2014, \apj, 796, 106,
  \dodoi{10.1088/0004-637X/796/2/106}

\bibitem[{{Jiang} {et~al.}(2019){Jiang}, {Stone}, \& {Davis}}]{Jiang:2019}
---. 2019, \apj, 880, 67, \dodoi{10.3847/1538-4357/ab29ff}

\bibitem[{{Katz} {et~al.}(2015){Katz}, {Sijacki}, \& {Haehnelt}}]{Katz:2015}
{Katz}, H., {Sijacki}, D., \& {Haehnelt}, M.~G. 2015, \mnras, 451, 2352,
  \dodoi{10.1093/mnras/stv1048}

\bibitem[{{Lupi} {et~al.}(2014){Lupi}, {Colpi}, {Devecchi}, {Galanti}, \&
  {Volonteri}}]{Lupi:2014}
{Lupi}, A., {Colpi}, M., {Devecchi}, B., {Galanti}, G., \& {Volonteri}, M.
  2014, \mnras, 442, 3616, \dodoi{10.1093/mnras/stu1120}

\bibitem[{{Madau} \& {Rees}(2001)}]{MadauR:01}
{Madau}, P., \& {Rees}, M.~J. 2001, \apjl, 551, L27, \dodoi{10.1086/319848}

\bibitem[{{Milosavljevi{\'c}} {et~al.}(2009){Milosavljevi{\'c}}, {Couch}, \&
  {Bromm}}]{MiloCB:09}
{Milosavljevi{\'c}}, M., {Couch}, S.~M., \& {Bromm}, V. 2009, \apjl, 696, L146,
  \dodoi{10.1088/0004-637X/696/2/L146}

\bibitem[{{Mortlock} {et~al.}(2011){Mortlock}, {Warren}, {Venemans}, {Patel},
  {Hewett}, {McMahon}, {Simpson}, {Theuns}, {Gonz{\'a}les-Solares}, {Adamson},
  {Dye}, {Hambly}, {Hirst}, {Irwin}, {Kuiper}, {Lawrence}, \&
  {R{\"o}ttgering}}]{Mortlock:2011}
{Mortlock}, D.~J., {Warren}, S.~J., {Venemans}, B.~P., {et~al.} 2011, \nat,
  474, 616, \dodoi{10.1038/nature10159}

\bibitem[{{Pacucci} \& {Ferrara}(2015)}]{PacucciF:2015}
{Pacucci}, F., \& {Ferrara}, A. 2015, \mnras, 448, 104,
  \dodoi{10.1093/mnras/stv018}

\bibitem[{{Pacucci} {et~al.}(2015){Pacucci}, {Volonteri}, \&
  {Ferrara}}]{PacucciVF:2015}
{Pacucci}, F., {Volonteri}, M., \& {Ferrara}, A. 2015, \mnras, 452, 1922,
  \dodoi{10.1093/mnras/stv1465}

\bibitem[{{Park} \& {Ricotti}(2011)}]{ParkR:11}
{Park}, K., \& {Ricotti}, M. 2011, \apj, 739, 2,
  \dodoi{10.1088/0004-637X/739/1/2}

\bibitem[{{Park} \& {Ricotti}(2012)}]{ParkR:12}
---. 2012, \apj, 747, 9, \dodoi{10.1088/0004-637X/747/1/9}

\bibitem[{{Park} \& {Ricotti}(2013)}]{ParkR:13}
---. 2013, \apj, 767, 163, \dodoi{10.1088/0004-637X/767/2/163}

\bibitem[{{Park} {et~al.}(2014{\natexlab{a}}){Park}, {Ricotti}, {Di Matteo}, \&
  {Reynolds}}]{ParkRDR:14b}
{Park}, K., {Ricotti}, M., {Di Matteo}, T., \& {Reynolds}, C.~S.
  2014{\natexlab{a}}, \mnras, 445, 2325, \dodoi{10.1093/mnras/stu1929}

\bibitem[{{Park} {et~al.}(2014{\natexlab{b}}){Park}, {Ricotti}, {Di Matteo}, \&
  {Reynolds}}]{ParkRDR:14a}
---. 2014{\natexlab{b}}, \mnras, 437, 2856, \dodoi{10.1093/mnras/stt2096}

\bibitem[{{Park} {et~al.}(2016){Park}, {Ricotti}, {Natarajan},
  {Bogdanovi{\'c}}, \& {Wise}}]{Park:2016}
{Park}, K., {Ricotti}, M., {Natarajan}, P., {Bogdanovi{\'c}}, T., \& {Wise},
  J.~H. 2016, \apj, 818, 184, \dodoi{10.3847/0004-637X/818/2/184}

\bibitem[{{Park} {et~al.}(2017){Park}, {Wise}, \&
  {Bogdanovi{\'c}}}]{ParkWB:2017}
{Park}, K., {Wise}, J.~H., \& {Bogdanovi{\'c}}, T. 2017, \apj, 847, 70,
  \dodoi{10.3847/1538-4357/aa8729}

\bibitem[{{Proga} \& {Begelman}(2003{\natexlab{a}})}]{ProgaB:2003a}
{Proga}, D., \& {Begelman}, M.~C. 2003{\natexlab{a}}, \apj, 582, 69,
  \dodoi{10.1086/344537}

\bibitem[{{Proga} \& {Begelman}(2003{\natexlab{b}})}]{ProgaB:2003b}
---. 2003{\natexlab{b}}, \apj, 592, 767, \dodoi{10.1086/375773}

\bibitem[{{Regan} {et~al.}(2019){Regan}, {Downes}, {Volonteri}, {Beckmann},
  {Lupi}, {Trebitsch}, \& {Dubois}}]{Regan:2019}
{Regan}, J.~A., {Downes}, T.~P., {Volonteri}, M., {et~al.} 2019, \mnras, 486,
  3892, \dodoi{10.1093/mnras/stz1045}

\bibitem[{{Regan} {et~al.}(2017){Regan}, {Visbal}, {Wise}, {Haiman},
  {Johansson}, \& {Bryan}}]{Regan:2017}
{Regan}, J.~A., {Visbal}, E., {Wise}, J.~H., {et~al.} 2017, Nature Astronomy,
  1, 0075, \dodoi{10.1038/s41550-017-0075}

\bibitem[{{Reinoso} {et~al.}(2018){Reinoso}, {Schleicher}, {Fellhauer},
  {Klessen}, \& {Boekholt}}]{Reinoso:2018}
{Reinoso}, B., {Schleicher}, D.~R.~G., {Fellhauer}, M., {Klessen}, R.~S., \&
  {Boekholt}, T.~C.~N. 2018, \aap, 614, A14,
  \dodoi{10.1051/0004-6361/201732224}

\bibitem[{{Ricotti}(2014)}]{Ricotti:2014}
{Ricotti}, M. 2014, \mnras, 437, 371, \dodoi{10.1093/mnras/stt1898}

\bibitem[{{Sakurai} {et~al.}(2016){Sakurai}, {Inayoshi}, \&
  {Haiman}}]{Sakurai:2016}
{Sakurai}, Y., {Inayoshi}, K., \& {Haiman}, Z. 2016, \mnras, 461, 4496,
  \dodoi{10.1093/mnras/stw1652}

\bibitem[{{Shakura} \& {Sunyaev}(1973)}]{ShakuraS:73}
{Shakura}, N.~I., \& {Sunyaev}, R.~A. 1973, \aap, 24, 337

\bibitem[{{Sugimura} {et~al.}(2018){Sugimura}, {Hosokawa}, {Yajima},
  {Inayoshi}, \& {Omukai}}]{Sugimura:2018}
{Sugimura}, K., {Hosokawa}, T., {Yajima}, H., {Inayoshi}, K., \& {Omukai}, K.
  2018, \mnras, 478, 3961, \dodoi{10.1093/mnras/sty1298}

\bibitem[{{Sugimura} {et~al.}(2017){Sugimura}, {Hosokawa}, {Yajima}, \&
  {Omukai}}]{Sugimura:2016}
{Sugimura}, K., {Hosokawa}, T., {Yajima}, H., \& {Omukai}, K. 2017, \mnras,
  469, 62, \dodoi{10.1093/mnras/stx769}

\bibitem[{{Takeo} {et~al.}(2018){Takeo}, {Inayoshi}, {Ohsuga}, {Takahashi}, \&
  {Mineshige}}]{Takeo:2018}
{Takeo}, E., {Inayoshi}, K., {Ohsuga}, K., {Takahashi}, H.~R., \& {Mineshige},
  S. 2018, \mnras, 476, 673, \dodoi{10.1093/mnras/sty264}

\bibitem[{{Turk} {et~al.}(2011){Turk}, {Smith}, {Oishi}, {Skory}, {Skillman},
  {Abel}, \& {Norman}}]{Turk:2011}
{Turk}, M.~J., {Smith}, B.~D., {Oishi}, J.~S., {et~al.} 2011, \apjs, 192, 9,
  \dodoi{10.1088/0067-0049/192/1/9}

\bibitem[{{Willott} {et~al.}(2003){Willott}, {McLure}, \&
  {Jarvis}}]{Willott:2003}
{Willott}, C.~J., {McLure}, R.~J., \& {Jarvis}, M.~J. 2003, \apjl, 587, L15,
  \dodoi{10.1086/375126}

\bibitem[{{Wise} \& {Abel}(2011)}]{Wise:2011}
{Wise}, J.~H., \& {Abel}, T. 2011, \mnras, 414, 3458,
  \dodoi{10.1111/j.1365-2966.2011.18646.x}

\bibitem[{{Yue} {et~al.}(2014){Yue}, {Ferrara}, {Salvaterra}, {Xu}, \&
  {Chen}}]{YueFSXC:14}
{Yue}, B., {Ferrara}, A., {Salvaterra}, R., {Xu}, Y., \& {Chen}, X. 2014,
  \mnras, 440, 1263, \dodoi{10.1093/mnras/stu351}

\end{thebibliography}
\label{lastpage}

\end{document}